\documentclass[article]{llncs}
\usepackage[a4paper, total={6in, 8in}]{geometry} 
\usepackage{amsmath}
\usepackage{booktabs} 
\usepackage{caption} 
\usepackage{subcaption} 
\usepackage{graphicx}
\usepackage{pgfplots}
\usepackage[all]{nowidow}
\usepackage[utf8]{inputenc}
\usepackage{tikz}
\usetikzlibrary{er,positioning,bayesnet}
\usepackage{multicol}
\usepackage{algpseudocode,algorithm,algorithmicx}
\usepackage{hyperref}
\usepackage{amsfonts}
\usepackage[section]{placeins}
\usepackage[inline]{enumitem} 
\definecolor{blue}{HTML}{1F77B4}
\definecolor{orange}{HTML}{FF7F0E}
\definecolor{green}{HTML}{2CA02C}

\pgfplotsset{compat=1.14}

\begin{document}
\title{Thermodynamics\\in Stochastic Conway's Game of Life}
%
%
\author{Krzysztof Pomorski\inst{1A,2} \and
Dariusz Kotula\inst{1B}}
%
%
\institute{Cracow University of Technology\\
A: Faculty of Electrical and Computer Engineering\\
B: Faculty of Computer Science and Telecommunications\\ \and
Quantum Hardware Systems (\href{www.quantumhardwaresystems.com}{www.quantumhardwaresystems.com})}
\maketitle              
\begin{abstract}
Cellular automata can simulate many complex physical phenomena using the power of simple rules. The presented methodological platform expresses the concept of programmable matter in which Newton’s laws of motion are one of examples. Energy has been introduced as the equivalent of the "Game of Life" mass, which can be treated as first level of approximation. The temperature presence and propagation was calculated for various lattice topology and boundary conditions by using the Shannon entropy measure.
The conducted study provides strong evidence that despite not fulfillment the principle of mass and energy conservation, the entropy, mass distribution and temperatures approaches thermodynamic equilibrium. In addition, the described cellular automata system transits from positive to a negative temperatures that stabilizes and can be treated as a signature of system dynamical equilibrium. Furthermore, the system dynamics was presented in case of few species of cellular automata competing for maximum presence on given lattice with different boundary conditions.
\end{abstract}
\section{Introduction to Classical Conway's Game of Life (CCGoL)}
\begin{figure}[H]
\centering
\includegraphics[height=.3 \linewidth]{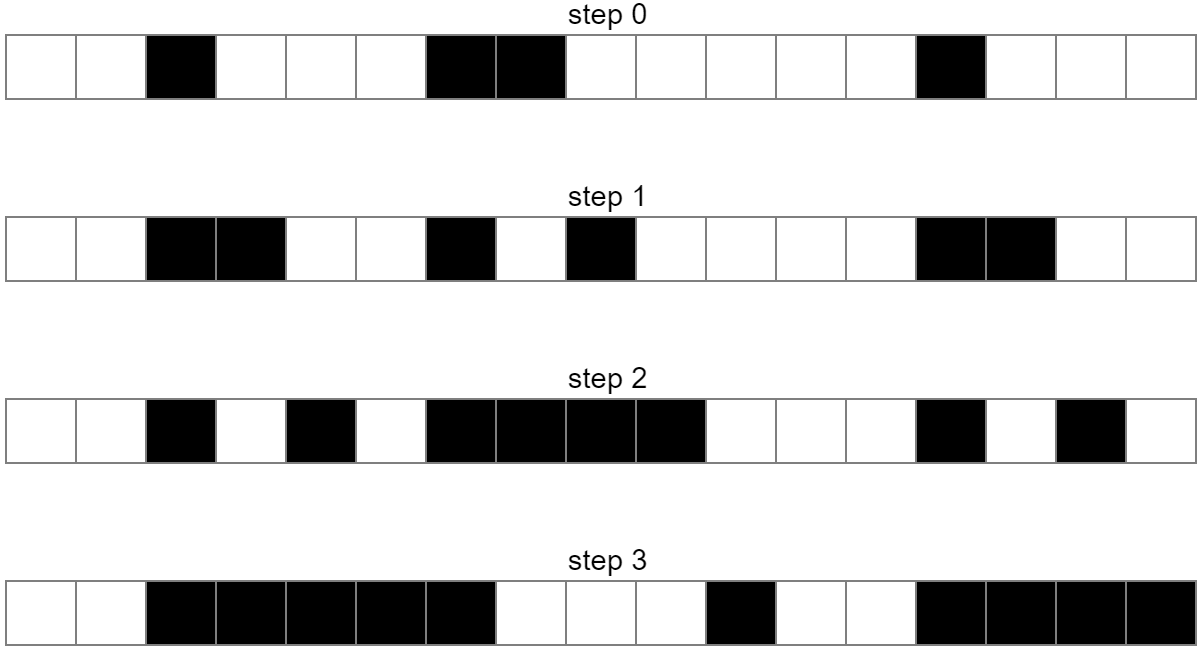}
\caption{Evolution of a one-dimensional cellular automaton in successive cycles with left side partial logical negation rule (if the state of nearest left cell is alive then given automata cell change its state to opposite).}
\label{fig:1D_automaton}
\end{figure}

\begin{figure}
\centering
\includegraphics[height=0.6 \linewidth]{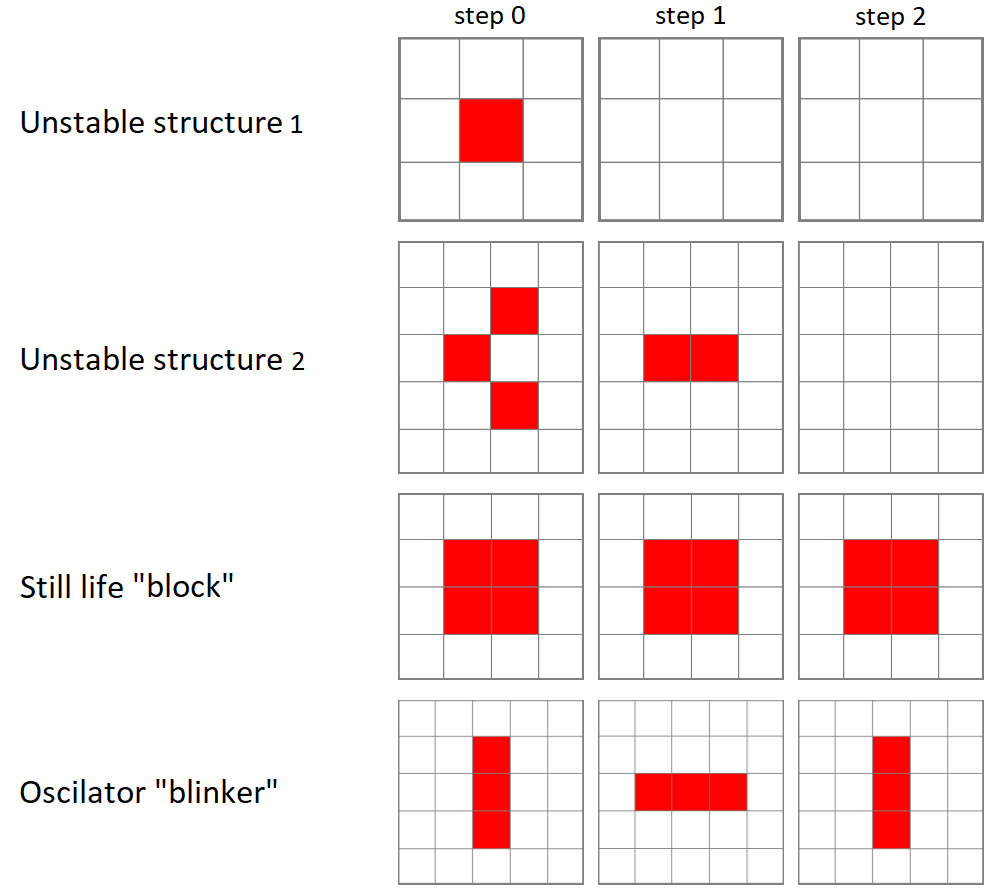}
\caption{Evolution of various topologies of cellular automata structures in time with deterministic rules of CCGoL. One can identify two dynamically unstable structures and two structures that have dynamical stability with time.}
\label{fig:structures}
\end{figure}

A cellular automaton is a system consisting of cells arranged most often on a one, two or three dimensional regular lattice, which at a given moment are in one of $M$-th possible states expressing M-valued logic. The dynamics of the model depends on the definition of individual cell states and the rules of transitions between them \cite{cite:4}. One of the simplest examples is a one-dimensional cellular automaton. Suppose that the cells placed on the lattice can be in one of two states, which are marked with white (default assigned to dead state or logical zero) or black color (default assigned to alive state or logical one). We define a rule that if a given cell is black, then the cell to the right of it will change its state. This situation is depicted in a Figure \ref{fig:1D_automaton}.
We can observe how the system changes in the subsequent steps of the simulation. The parameter determining the change of the cell state is the state of the left neighbour of a given cell. There are many other possibilities to choose such a parameter, e.g. the condition of the state of both neighboring cells or having the nearest neighbors with opposite states. In order to determine system dynamics, we must have a defined initial cellula automata states (information about initial system dynamical state) and a specific set of deterministic or probabilistic rules.
\noindent The Conway's Game of Life \cite{cite:1} is an example of a cellular automaton with a deterministic rules. It was proposed in 1970 by John Conway and cellular automaton system consists of cells located on a two-dimensional lattice, which can be in one of two states: alive or dead. The rules specify the required number of neighbors and cell states that are taken into account to determine their state in the next cycle. Given the neighborhood through the 8 closest cells, we can write the 3 main rules of the Classical Conway's Game of Life (CCGoL):
\begin{enumerate}
    \item If a dead cell has exactly 3 neighbors, it comes alive in the next cycle.
    \item If a living cell has 2 or 3 neighbors, it survives in the next cycle.
    \item If a cell has a different number of neighbors than stated above, it will be dead
in the next cycle.
\end{enumerate}
The rules defined in this way allow for the generation of various types of structure topologies with automaton state set to 1, as shown in Figure \ref{fig:structures}.
The most common are "unstable structures", which change in successive cycles, but do not return to their initial state. A single cell cannot survive on the lattice, because it has less than 2 neighbors. Dead cell surrounded by alife cells cannot come to life, because it has number of neighbors is different from 3. If the simulation is continuing sufficiently long time, on the lattice usually remain structures that are unchanging in time - "still lifes" (an example is the "block" shown in the Figure \ref{fig:structures}) or changing in time in periodic way, so they return to their original shape after $k$ cycles - "oscillators" (an example is the "blinker" shown in the Figure \ref{fig:structures}).
\begin{figure}
\centering
\subfloat[\centering]{{\includegraphics[height=0.15 \linewidth]{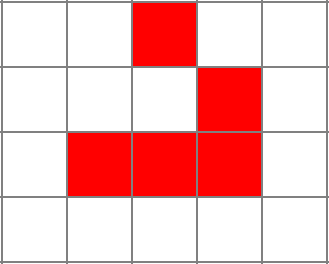} }}%
\subfloat[\centering]{{\includegraphics[height=0.15 \linewidth]{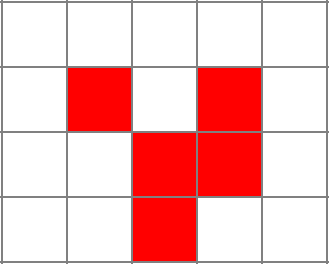} }}%
\subfloat[\centering]{{\includegraphics[height=0.15 \linewidth]{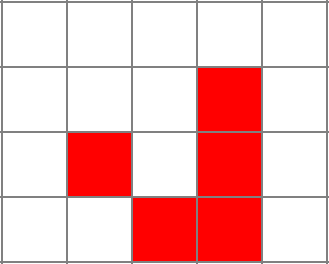} }}%
\subfloat[\centering]{{\includegraphics[height=0.15 \linewidth]{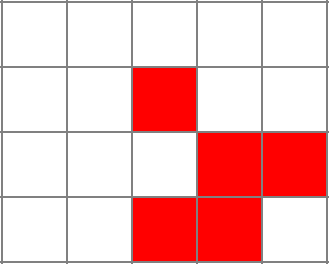} }}%
\subfloat[\centering]{{\includegraphics[height=0.15 \linewidth]{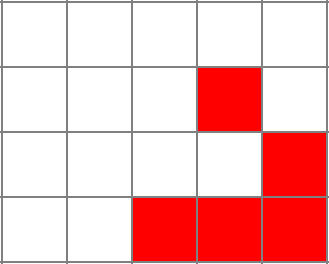} }}%
\caption{Evolution of the "glider" configuration of cellular automata having propagation property with time in deterministic CCGoL.}
\label{fig:glider}
\centering
\includegraphics[height=.3 \linewidth]{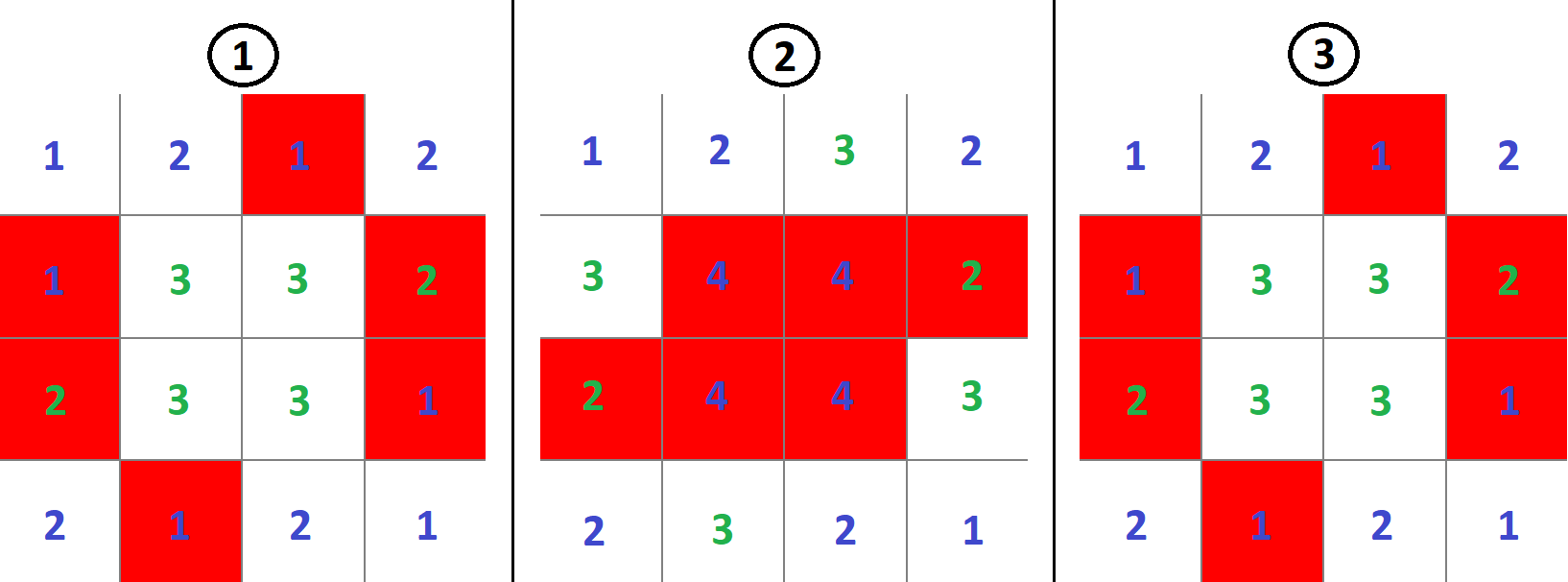}
\caption{Evolution of the "toad" configuration of cellular automata being an oscillator in successive cycles in CCGoL.}
\label{fig:oscilator}
\end{figure}
There are also structures that move in a certain direction - "gliders" (Figure \ref{fig:glider}) that are analogical to 1st Newton dynamics preserving momentum (speed and direction of propagation in this case), leave a trace of "blinkers" - "star ships" and objects, which periodically eject "gliders" - "guns" \cite{cite:5}.
Figure \ref{fig:oscilator} shows one among many existing in the Conway's Game of Life oscillators during successive iterations of the system simulation.

The "toad" oscillator has a period of 2, which means that it has continuous switching between 2 different fixed configurations. Each 2-dimensional discrete lattice field has a specified number that indicates the number of neighbors of the given cell. If the number is green, the cell will be alive in the next cycle. If the number is blue, that cell will be dead in the next cycle.

\section{Introduction to Stochastic Classical Conway's Game of Life (SCCGoL)}
\begin{figure}
\centering
\subfloat[\centering]{{\includegraphics[height=4cm]{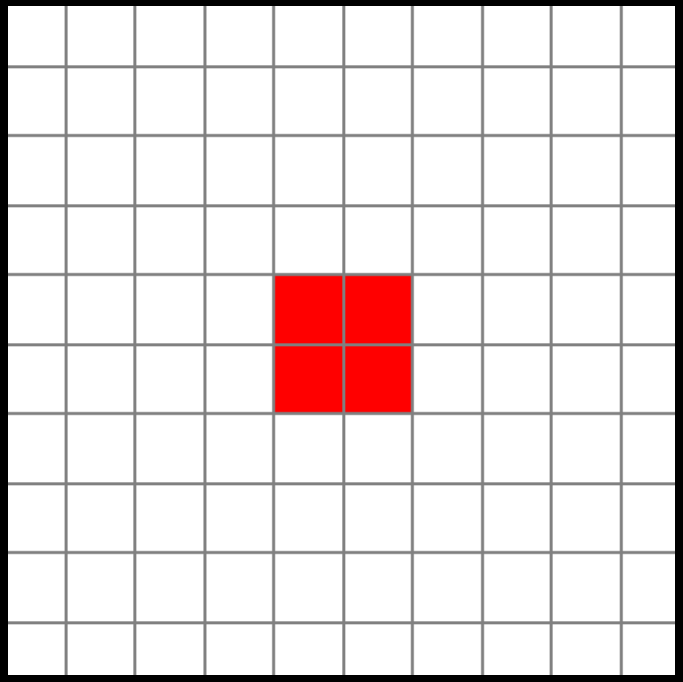} }}%
\qquad
\subfloat[\centering]{{\includegraphics[height=4cm]{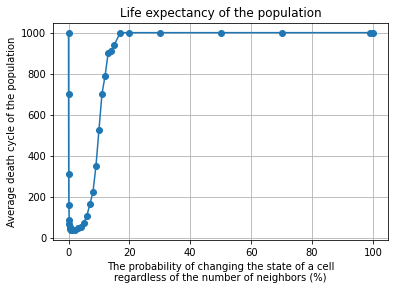} }}%
\qquad
\subfloat[\centering]{{\includegraphics[height=4cm]{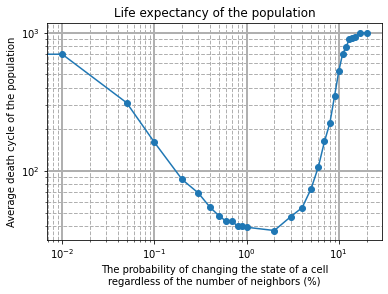} }}%
\caption{(a) Schematic view of initial conditions in SCCGoL. (b-c) Dependence of automata population average cycle lifetime (over 1000 trials) on probability of spontaneous change of cell
state from life to dead and reversely with preservation of standard rules in Conway's Game of Life.}
\label{fig:life_expectancy}
\end{figure}

The Stochastic Classical Conway's Game of Life (SCCGoL) was created by an addition of a cell spontaneous change probability to the rules that were initially deterministic, so stochastic determinism was achieved. With a given prefixed spontaneous probability value, the state of the cell can change regardless of the number of neighbors. Cell states have values between 0 and 1 and are called mass. Due to the fact that SCCGoL have different rules from CCGoL, cells have almost never exactly two or three neighbors. A condition for given cell to come to life from a dead cell state (creationism of alife cell) is by having a number of neighbors in a certain range of values. Similar rules applies to a living cell justifying its alife or dead state in next time iteration. By setting standard intervals of allowed/forbidden number of neighbours values, in which cell is alife/dead and by adding the additional spontaneous rule probability for cell to change in the next iteration (probability of
changing the state of a cell regardless of the number of neighbors), we are able to create a simulator, where the cells never die or it is very difficult from a probabilistic point of view for a given cell to stay alive. If, maintaining standard neighbours condition for being dead/alife in previously defined intervals and having a selected initial cell alife automata configuration, we can systematically increase the spontaneous probability level from 0\% to 100\%, and we result in the graph as depicted in Figure \ref{fig:life_expectancy}.
Simulations were conducted for a lattice size 10 by 10 with the initial condition of a cellular automaton 2 by 2 (Figure \ref{fig:life_expectancy}a) with limited maximum number of cycles set to 1000 in conducted simulations. For a probability of changing the state of a cell regardless of the number of neighbors equal to zero, there is full determinism, therefore it is a situation of CCGoL with changed numbers of neighbors. As the probability increases (in range from 0 to 2\%), the life expectancy of the population decreases due to too few neighbors. Starting simulation from a probability level close to two percent (probability of changing the state of a cell regardless of the number of neighbors), the average life expectancy of the population begins to rise, which is caused by the more frequent appearances of living cells. Conducting the simulation with a probability level above eighteen percent we observe that the population practically never dies and keeps average cycle life time being at least 1000 or more time iterations.

\section{Generalization of Stochastic Classical Conway’s Game of Life (SCCGoL) to the case of N competing cellular automata species}
\begin{figure}
\centering
\subfloat[\centering]{{\includegraphics[width=.3 \linewidth]{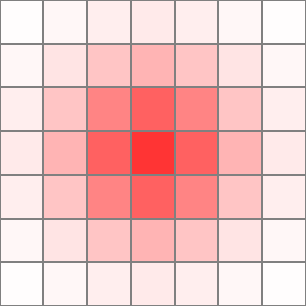} }}%
\qquad
\subfloat[\centering]{{\includegraphics[width=.3 \linewidth]{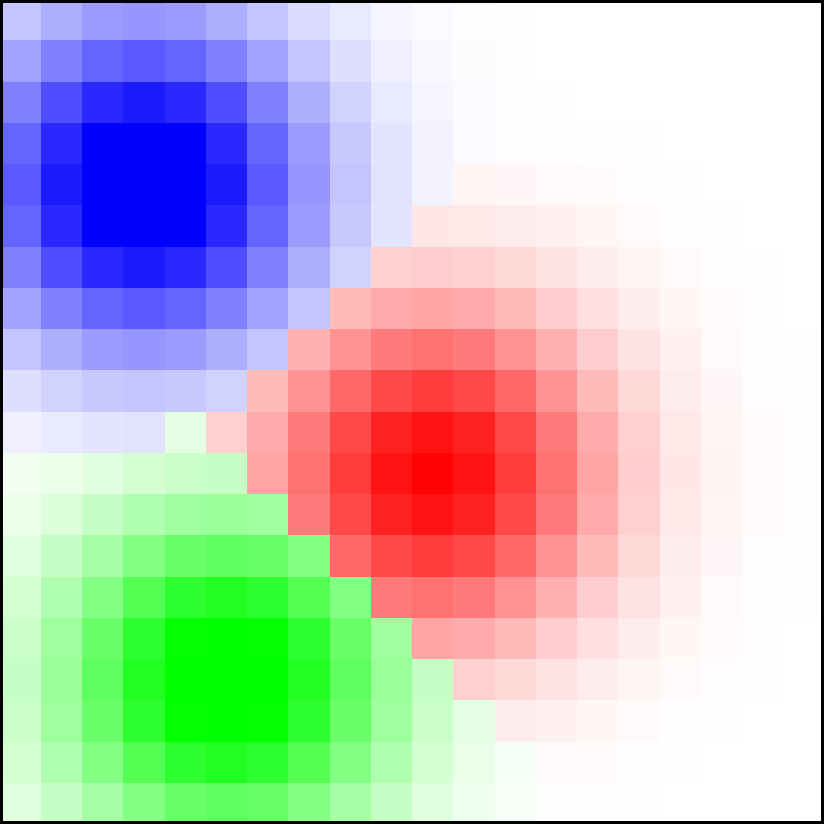} }}%
\caption{(a) Case of mass distribution of cellular automata of one species using a two-dimensional Gaussian function being isotropic. (b) Example of initial distribution of three cellular automata tribes with the case of boundary existence between each separate tribe with a rule that one geometrical place on the lattice is occupied by cellular automata of given species with dominant mass.}
\label{fig:species}
\end{figure}

\begin{figure}
\centering
\subfloat[\centering]{{\includegraphics[width=.3\linewidth]{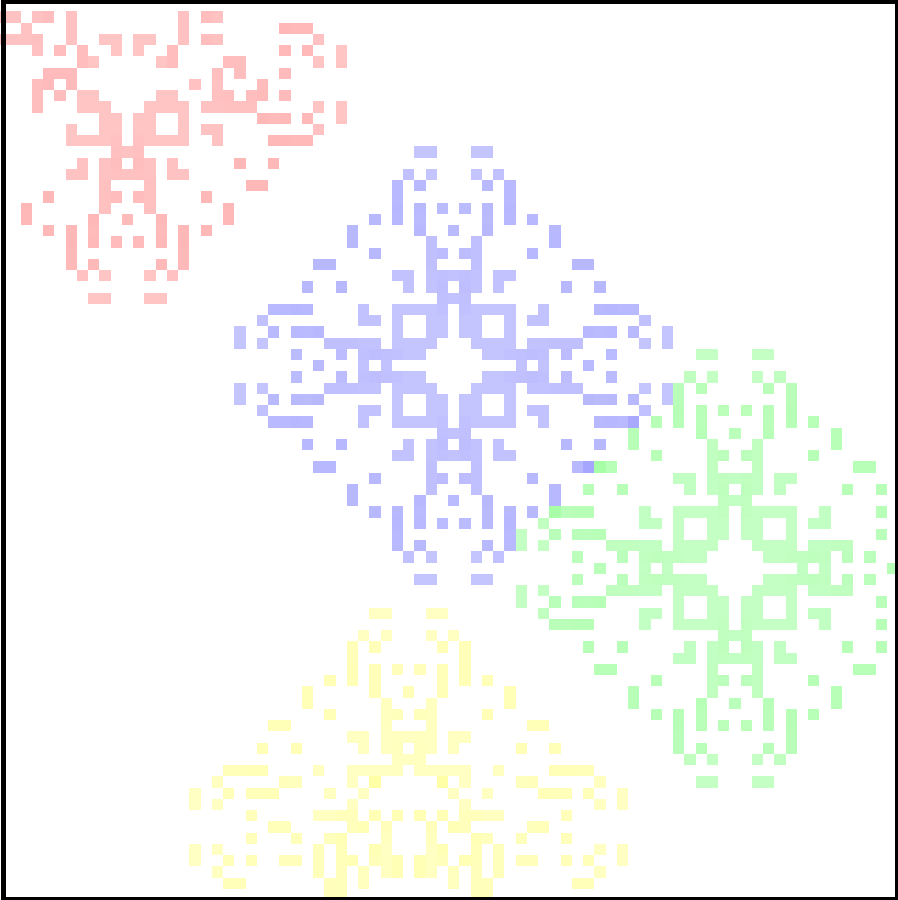} }}%
\qquad
\subfloat[\centering]{{\includegraphics[width=.3\linewidth]{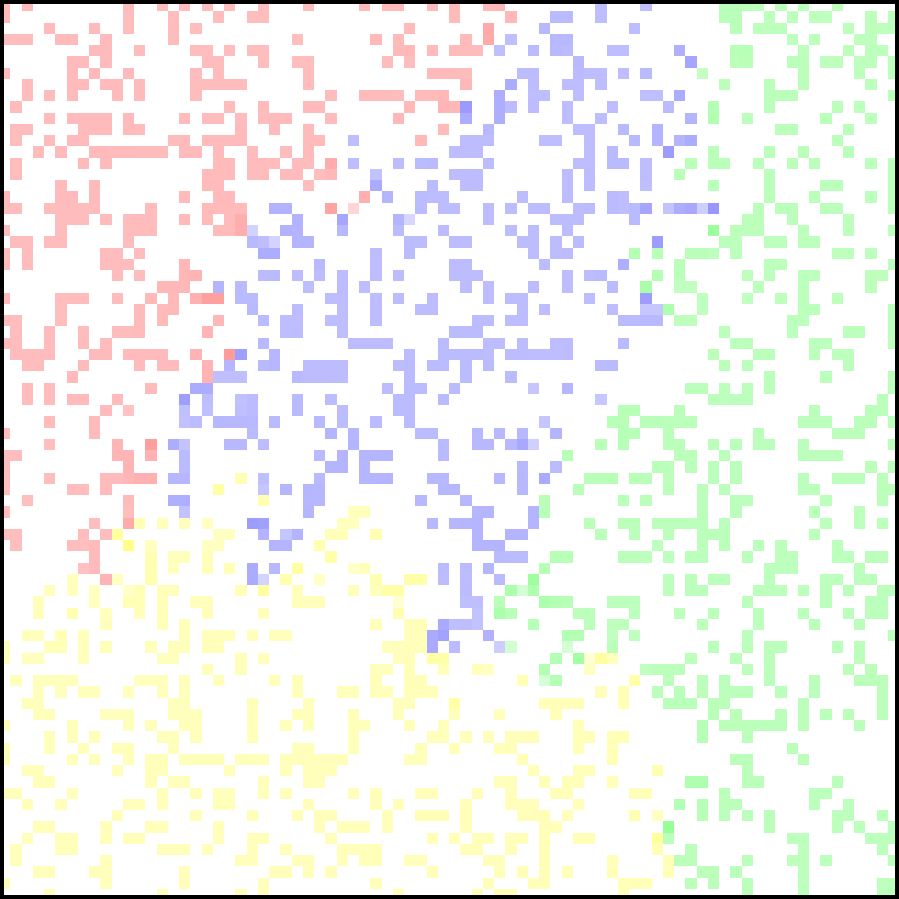} }}%
\caption{Evolution of map distribution of 4 cellular automata populations of different species with 100 by 100 lattice size. One can spot effective tendency of each species to occupy maximum possibly territory at the cost of other species in 4-species SCCGoL (SCCGoL4S) what can be understand as occurrence of weak antagonistic relation}
\label{fig:species2}
\end{figure}

The created simulation platform enables to create $N$ different cellular automata species that competes between themselves for the resources enabling them to replicate. The given automata species has better replication properties within its own community and worse replication properties in case of neighbors being of different species.
Such situation is normally encountered in human population when people of one homogenous identity prefer to collaborate more than in the case of people having much different identity (being from different tribes). Due to that fact soft antagonistic relation between different species is introduced indirectly by means of higher level of tolerance (or effectiveness of replication) towards neighbors of own species than neighbors of other species (other different species are treated in the same way). In real way this is an analogy of  members collaboration of a given nation or culture within given culture or community versus different culture or community. Let us consider $N=4$ (number of different automata species), so we have the following determined formula for number of effective existing neighbors $Ne_{effective,k}$ for given k-th tribes as:
\begin{equation}
Ne_{effective,k}=a_{1,k}Ne_1+a_{2,k}Ne_2+...+a_{k,k}Ne_k+...+a_{k,N}Ne_N.
\label{Neff}
\end{equation}
Previously defined replication rules promoting its own species can be formally expressed by following condition (max($a_{s,k})=a_{k,k}$ and $a_{k,k}> a_{s,k}$ if $s\neq k$). In the conducted simulations all tribes have assigned the same value ($a_{1,1}=a_{2,2}=a_{3,3}=a_{4,4}=\frac{1}{2}$) and another same value for ($a_{1,2}=a_{1,3}=a_{1,4}=a_{2,1}=a_{2,3}=a_{2,4}=a_{3,1}=a_{3,2}=a_{3,4}=a_{4,1}=a_{4,2}=a_{4,3}=\frac{1}{3}$) what simply means that automata tribes promotes its own tribe and only partly promotes other tribes with no special distinction on which different tribe it is pointing to.
Cellular automata species distribution for k-th tribe across two-dimensional lattice is generated with use of a two-dimensional Gaussian distribution given as follows:
\begin{equation}
f(x,y)_k=A_ke^{-\left(\frac{(x-x_{0,k})^2}{2\sigma_{x,k}^2}+\frac{(y-y_{0,k})^2}{2\sigma_{y,k}^2}\right)},
\end{equation}
where $f(x,y)_k$ is a mass function depending on the cell coordinates, $(x_{0,k};y_{0,k})$ the center of the given cellular automaton species. Given parameters $A_{0,k}$ and $\sigma_{x,k}$, $\sigma_{y,k}$ can control the mass of species and spread across the "Globe" as depicted in Figure \ref{fig:species}.
In the case of several species on the lattice, SCCGoL simulation results promote the formation of new cells among the species that has the most mass in the neighborhood as in accordance in the Figure \ref{fig:species2}. During life cycle always newly formed cell has a mass that is more dependent on cells of the same species than other species of its neighbors. We observed that automata species fight each other in order to spread across the lattice and exclude other species what is an expression of some form of biological Darwinism being a consequence of formula \ref{Neff}.

\section{Methodology of description of SCCGoL dynamics by tools of classical statistical physics}
\begin{figure}
\centering
\includegraphics[height=.3 \linewidth]{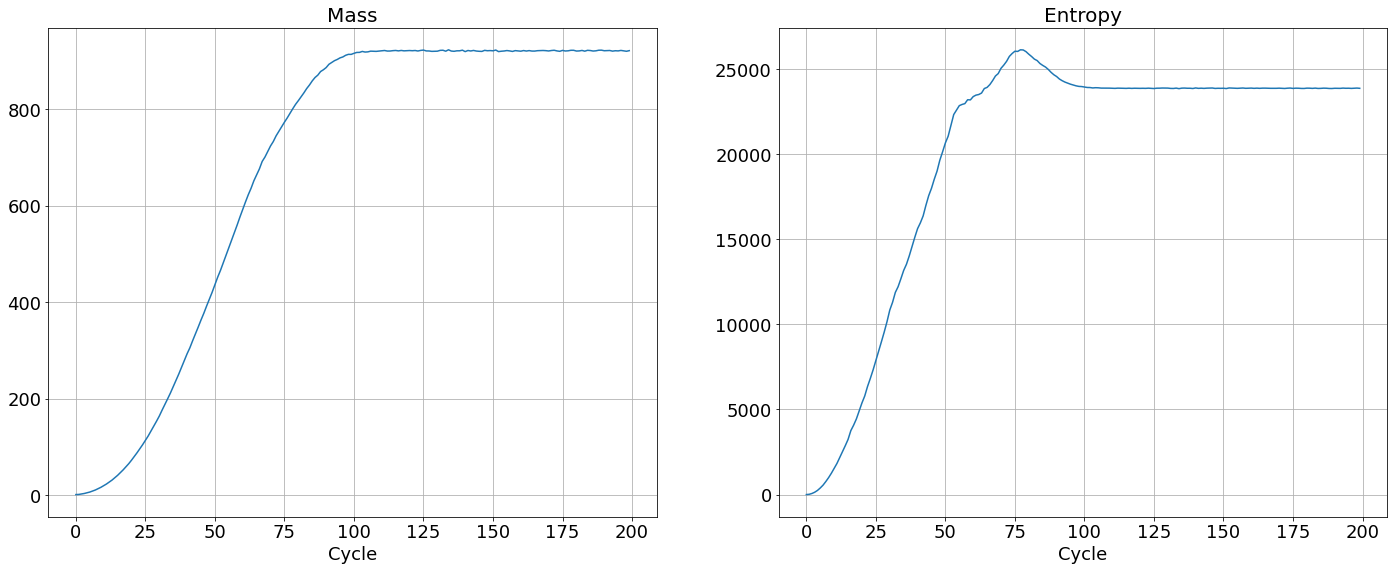}
\caption{Evolution of mass and entropy in successive cycles in SCCGoL manifesting maximization with saturation and approaching stationary thermodynamical equilibrium with initial condition depicted in Figure \ref{fig:no_barriers_1000}.}
\label{fig:mass_entropy_no_barriers_1000}
\end{figure}

Adding probability to the initially deterministic Game of Life and changing the interaction rules between neighboring automata brought the necessity for a new quantity called mass that has continuous real values (other than 0 and 1 present in deterministic Game of Life).
At this stage vividness of the whole cellular automata population can be understood and approximated by the population total energy (where simply mass is equal to energy). The level of the automata distribution order is expressed by the entropy. Entropy was introduced by Rudolf Clausius in 1865 as a thermodynamic state function \cite{cite:3}. If the entropy at the initial state and the entropy of the final state are devoted by $S_i$ and $S_f$ respectively, then we have:
\begin{equation} \label{eq:thermodynamic_entropy}
S_f-S_i=\int_i^f\frac{dE}{T}, \frac{dS}{dE}=\frac{1}{T}
\end{equation}
Very last formula gives definition of a temperature under assumption that energy and entropy is defined that is the case. We use instead of thermodynamic entropy the Shannon entropy measure given as:
\begin{equation} \label{eq:shannon_entropy}
S_{Shannon}=-\sum_{i,j} p(x_i,y_j)\log{p(x_i,y_j)}=\mathbb{E}\left[-\log{p(x,y)}\right]
\end{equation}
In order to calculate the probability in the equation \ref{eq:shannon_entropy}, the SCCGoL simulation is repeated several hundred times, so one determines an average population cell mass in successive cycles.
The calculation of the temperature, which is a measure of thermal state was achieved with an equation \ref{eq:thermodynamic_entropy} with a changed form:
\begin{equation}
T(x,y,t)=\frac{\frac{dE(x,y,t)}{dt}}{\frac{dS_{Shannon}(x,y,t)}{dt}}=\frac{dE(x,y,t)}{dS_{Shannon}(x,y,t)}
\end{equation}
The temperature defined by last formula can be calculated by two different methods. The first method relies on the calculation of the energy and entropy for the entire system, and then numerical calculation of the derivative of energy in function of entropy. The second calculation approach is to differentiate mass and entropy with respect to simulation time for each cell and as a result of corresponding ratio we obtain a temperature map. The example of evolution of mass and entropy for SCCGoL is depicted in Figure \ref{fig:mass_entropy_no_barriers_1000} and shows maximization and saturation of entropy, what is confirmation of Second Thermodynamic Law that is valid in physical systems, while we are dealing with cellular automata system. On another hand the mass of a system saturates and increases to certain critical value as given in left part of Figure \ref{fig:mass_entropy_no_barriers_1000}.


\section{Numerical analysis of SCCGoL dynamics with methodology of classical statistical physics}
\begin{figure}
\centering
\subfloat[\centering]{{\includegraphics[width=.3\linewidth]{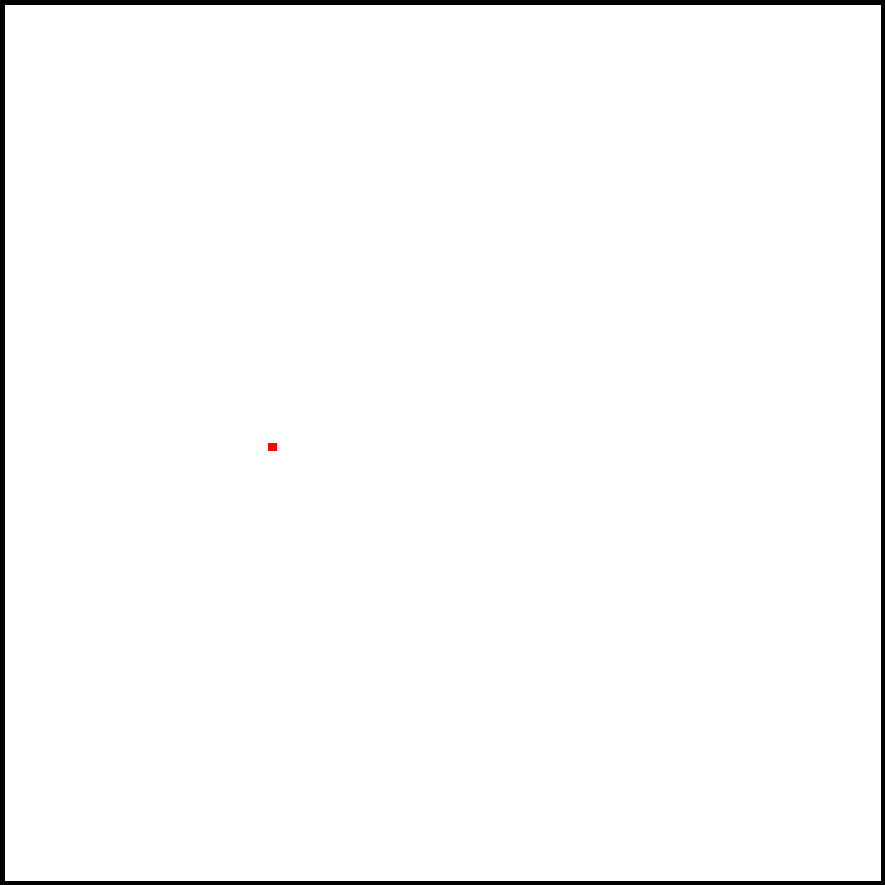} }}%
\qquad
\subfloat[\centering]{{\includegraphics[width=.3\linewidth]{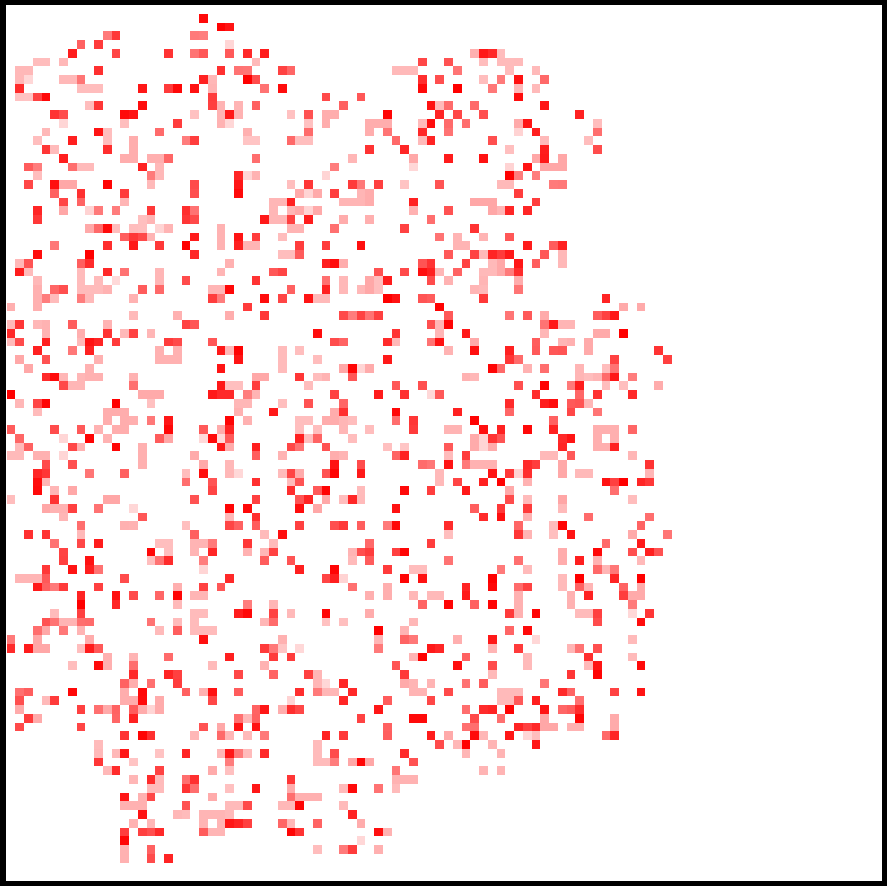} }}%
\caption{Evolution of diffusion process in cellular automata system for a limited lattice size 100 by 100 in SCCGoL (averaged over 1000 trials). It shows final saturation of mass and entropy (as depicted in Figure \ref{fig:mass_entropy_no_barriers_1000}) what implies approaching thermodynamical equilibrium with characteristic fluctuations of mass and entropy around effective stationary values.}
\label{fig:no_barriers_1000}
\end{figure}

\begin{figure}
\centering
\subfloat[Mass at t=4 \centering]{{\includegraphics[width=.3\linewidth]{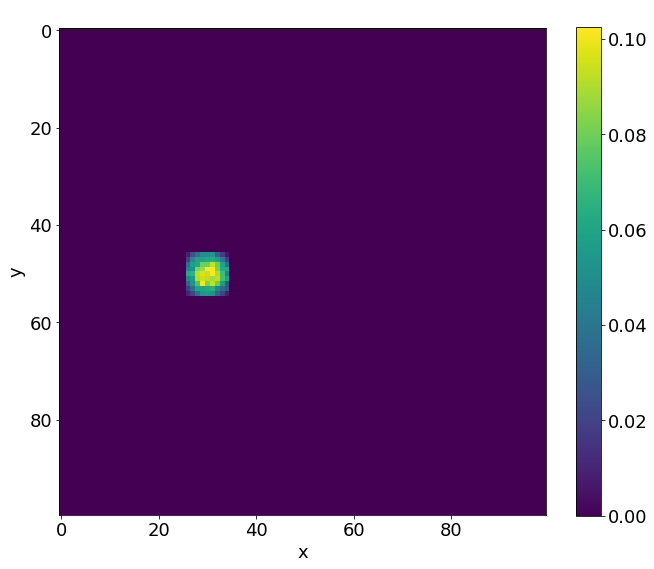}}}%
\subfloat[Mass at t=26 \centering]{{\includegraphics[width=.3\linewidth]{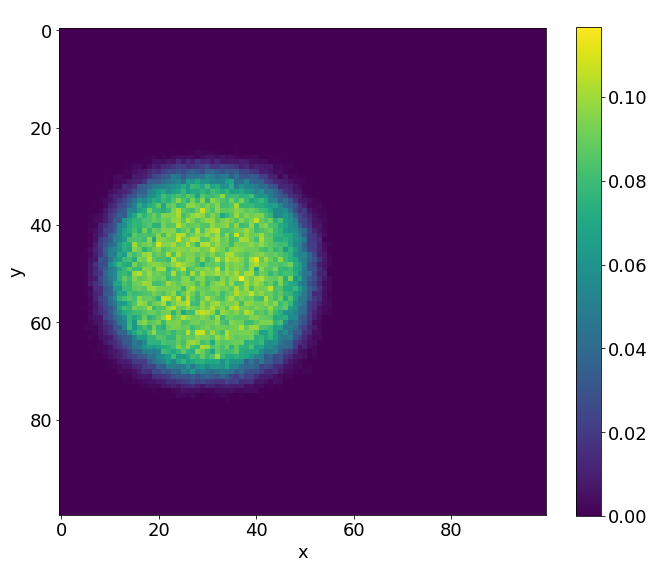} }}%
\subfloat[Mass at t=92 \centering]{{\includegraphics[width=.3\linewidth]{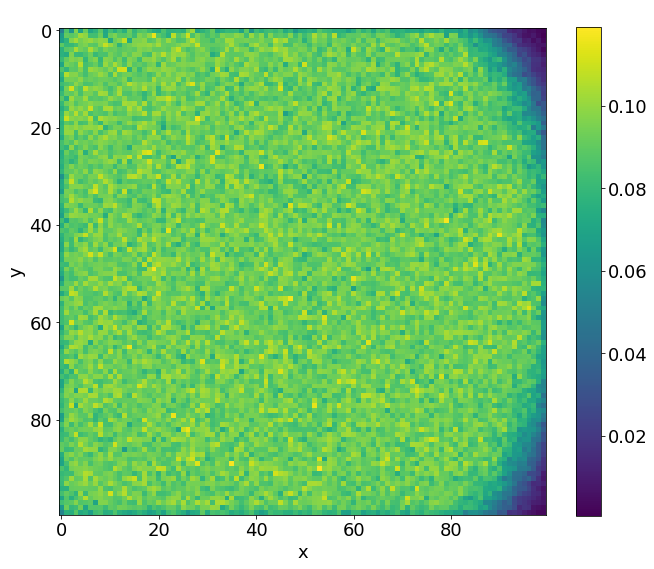} }}%
\qquad
\subfloat[Entropy at t=4 \centering]{{\includegraphics[width=.3\linewidth]{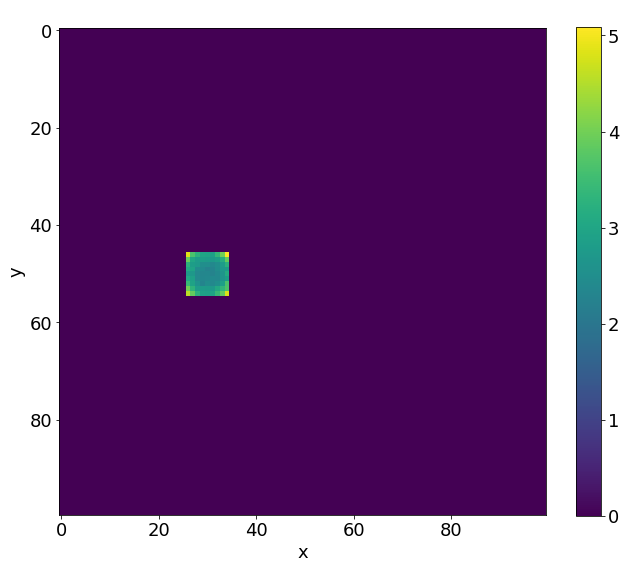} }}%
\subfloat[Entropy at t=26 \centering]{{\includegraphics[width=.3\linewidth]{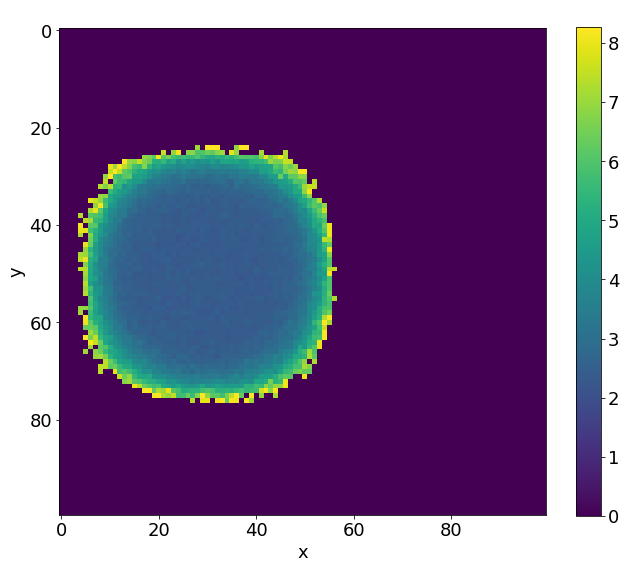} }}%
\subfloat[Entropy at t=92 \centering]{{\includegraphics[width=.3\linewidth]{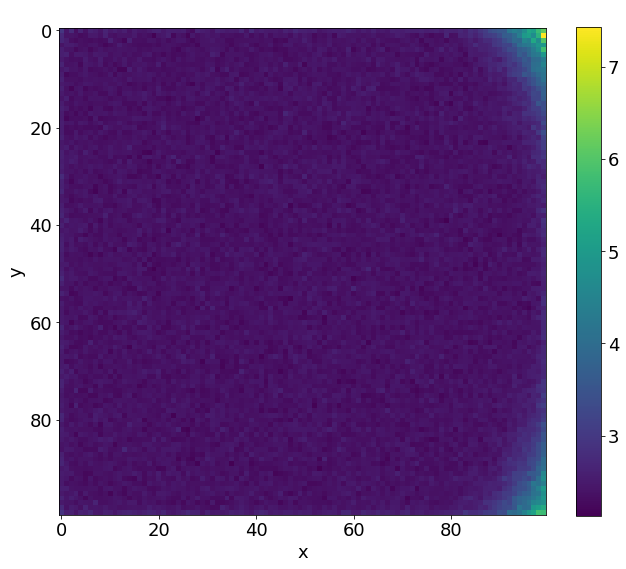} }}%
\qquad
\subfloat[T(x,y,t=4) \centering]{{\includegraphics[width=.3\linewidth]{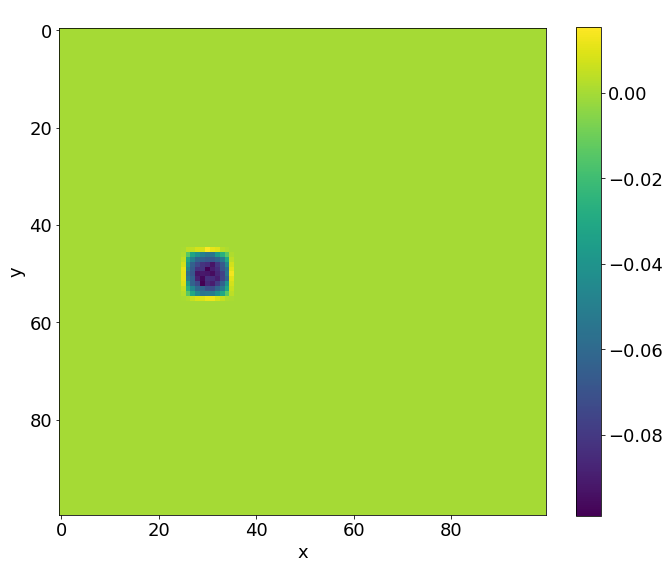} }}%
\subfloat[T(x,y,t=26) \centering]{{\includegraphics[width=.3\linewidth]{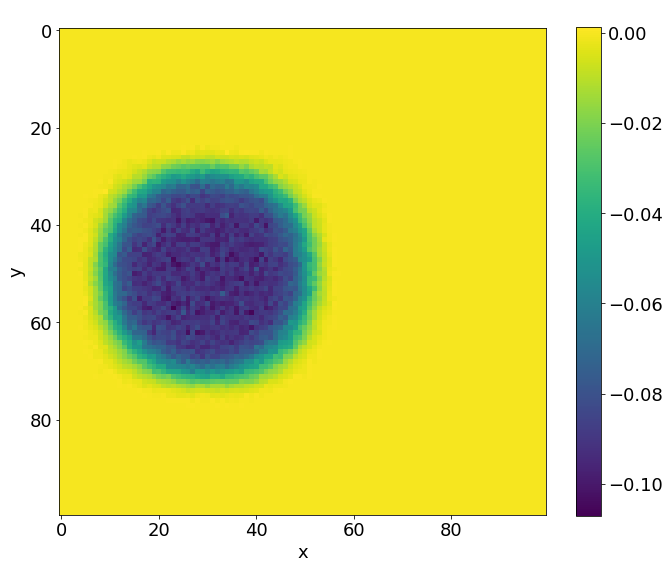} }}%
\subfloat[T(x,y,t=92) \centering]{{\includegraphics[width=.3\linewidth]{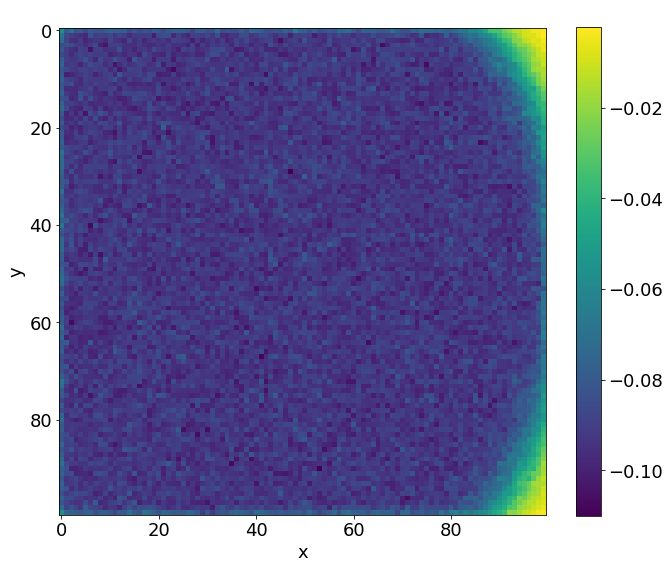} }}%
\qquad
\subfloat[$\frac{dm}{dt}$ with time \centering]{{\includegraphics[height=.215\linewidth]{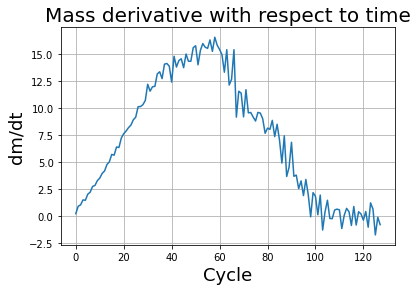} }}%
\subfloat[$\frac{dS}{dt}$ with time \centering]{{\includegraphics[height=.215\linewidth]{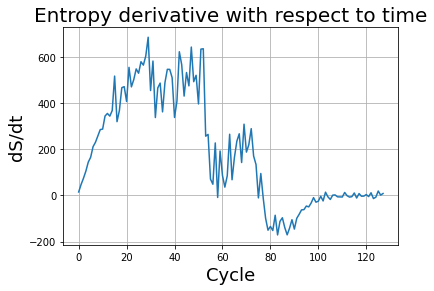} }}%
\subfloat[Temperature with time\centering]{{\includegraphics[height=.215\linewidth]{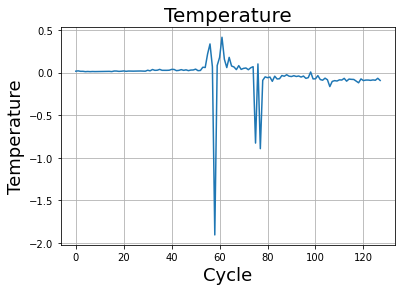} }}%
\caption{Dynamics of thermodynamical parameters (mass, entropy and temperature) with simulation time in SCCGoL (lattice size 100 by 100) also given in Figure \ref{fig:mass_entropy_no_barriers_1000}. Achieved thermodynamical equilibrium is accompanied with final distribution of negative temperature (starting from positive temperature distribution as in accordance with formula \ref{eq:thermodynamic_entropy}) as experimentally observed necessary criteria for final thermodynamical stability. One can spot various similarities of statistical behaviour of SCCGoL with physical systems described by classical thermodynamics (maximization and saturation of entropy, decay of temperature gradients and final thermalization, uniform distribution of mass and energy).}
\label{fig:characteristics_no_barriers_1000}
\end{figure}

Numerical simulations were carried out for 4 topologies cellular automata (case of Figures \ref{fig:no_barriers_1000}a, \ref{fig:one_barrier}a, \ref{fig:two_barriers}a, \ref{fig:two_tribes}a). We preinpose such a rule that given alive cell has 20\% probability of changing its state to dead state and that initially dead cell has 20\% probability of changing its state to alive with a mass choose randomly from an interval value in (0,0.5). Still given cell state is depended on its neighbors, since it has 80\% probability of changing its state due to state of neighbors. In order to obtain cellular automata probability map dynamics, simulations average of cellular automata positions was conducted. Figure \ref{fig:characteristics_no_barriers_1000} shows results that were obtained for a lattice size 100 by 100 with a one cell alive as initial lattice state (case of Figure \ref{fig:no_barriers_1000}a).
With subsequent cycles, the cells occupy more and more space on the lattice, which can be seen as increase the mass of the entire system (possible mass creationism is inherent feature of Conway's Game of Life). The sum of the masses of all cells in successive cycles is depicted in Figure \ref{fig:mass_entropy_no_barriers_1000} with a comparison of the entropy change of the entire system. As we observe in Figure \ref{fig:characteristics_no_barriers_1000}e, high entropy occurs at the edges of the population, which is caused by the entropy wave that meets area with almost zero cell occurrence.
Before equilibrium is established, the entropy of the system slightly decreases due to the loss of this extra entropy at the edges as can be seen in the right part of Figure \ref{fig:mass_entropy_no_barriers_1000}. Having established those two quantities, we conduct their differentiating with respect to time, and by formula \ref{eq:thermodynamic_entropy} one can establish the whole effective temperature of system by dividing change of mass at given time by change of the entropy.
Figure \ref{fig:characteristics_no_barriers_1000}k shows the greater susceptibility of the entropy change to the constraints associated with the limited size of the simulation lattice that is ended with impenetrable walls. The dependence of mass derivative with respect to time simulation (case of Figure \ref{fig:characteristics_no_barriers_1000}j) does not have such large oscillations as in the case of entropy derivative with simulation simulation (case of Figure \ref{fig:characteristics_no_barriers_1000}k). The temperature calculated by such procedure gives values just above zero (slightly positive) up to the 50'th cycle.  We can see a large down peak caused by a slowdown in entropy increase. From about the 75'th cycle, the temperature of the system goes from positive to negative values. A Figure \ref{fig:characteristics_no_barriers_1000}l) describes anomalous thermalization process in SCCGoL with a case of approaching temperature and entropy equilibrium. Surprisingly, the thermodynamic equilibrium is achieved for a case for negative temperatures. A second possible approach in determination the temperature is by use the derivatives of the mass and entropy with respect to time of the individual cells and by obtaining a temperature map of all cells.
The temperature depicted in Figure \ref{fig:characteristics_no_barriers_1000}i is mostly negative and steady, what corresponds to a situation where mass and entropy have come to equilibrium. We can distinguish two regions in the simulation, the first one with no temperature gradient and zero negative temperature and the second region with non-zero temperature gradient that also include positive temperatures. The place, where time derivatives of mass and entropy have noticeable values is at the edges of automata population, where we observe slightly positive temperature, which corresponds to the situation in Figure \ref{fig:characteristics_no_barriers_1000}l before the 75'th cycle.
\section{Numerical study of one species cellular automata with various boundary conditions}
Next family of simulations were conducted with use of barriers impenetrable by cellular automata cells via which cellular automata cannot interact with each other.
\begin{figure}
\centering
\subfloat[\centering]{{\includegraphics[width=.3 \linewidth]{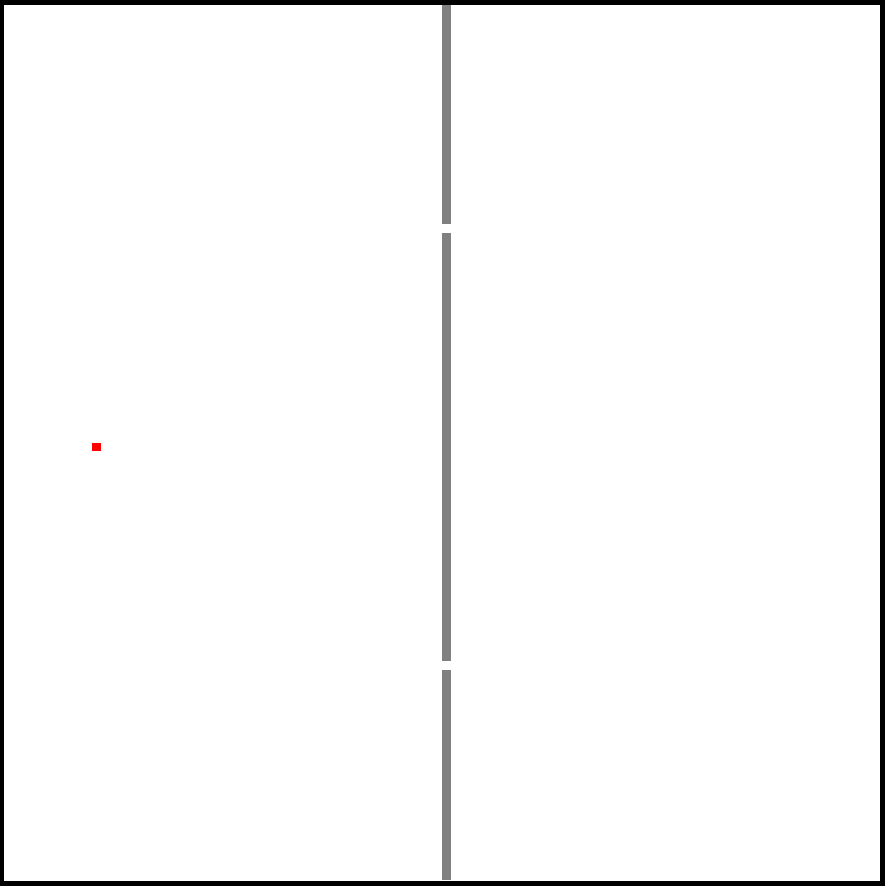} }}%
\qquad\qquad
\subfloat[\centering]{{\includegraphics[width=.3 \linewidth]{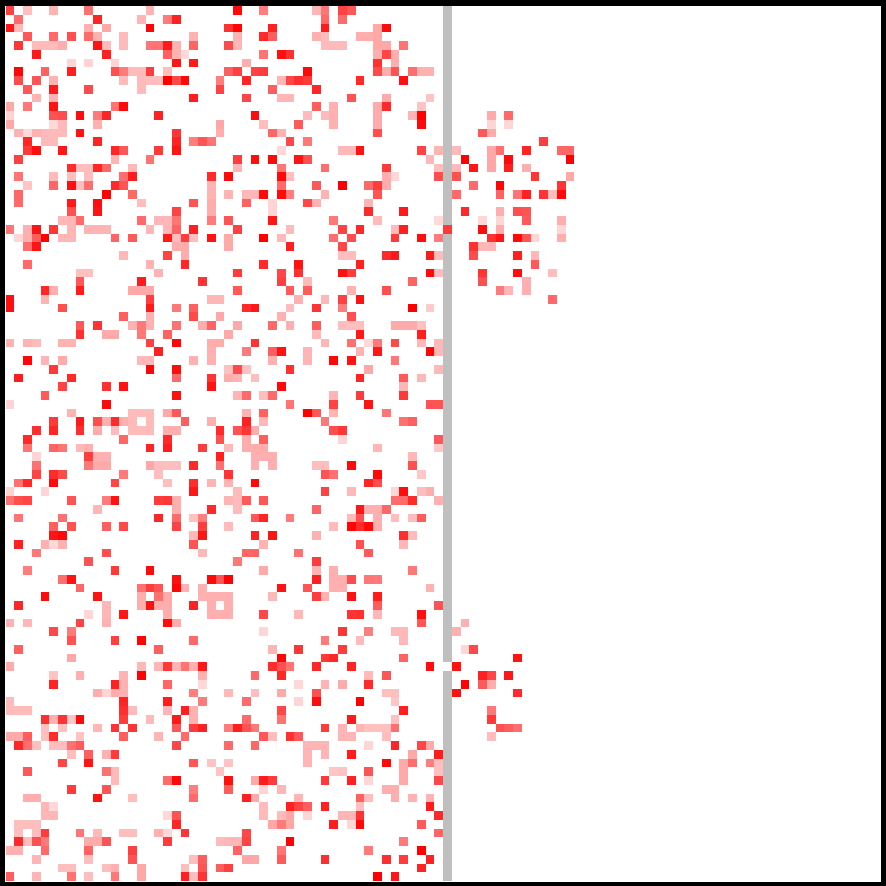} }}%
\qquad
\subfloat[\centering]{{\includegraphics[height=.3 \linewidth]{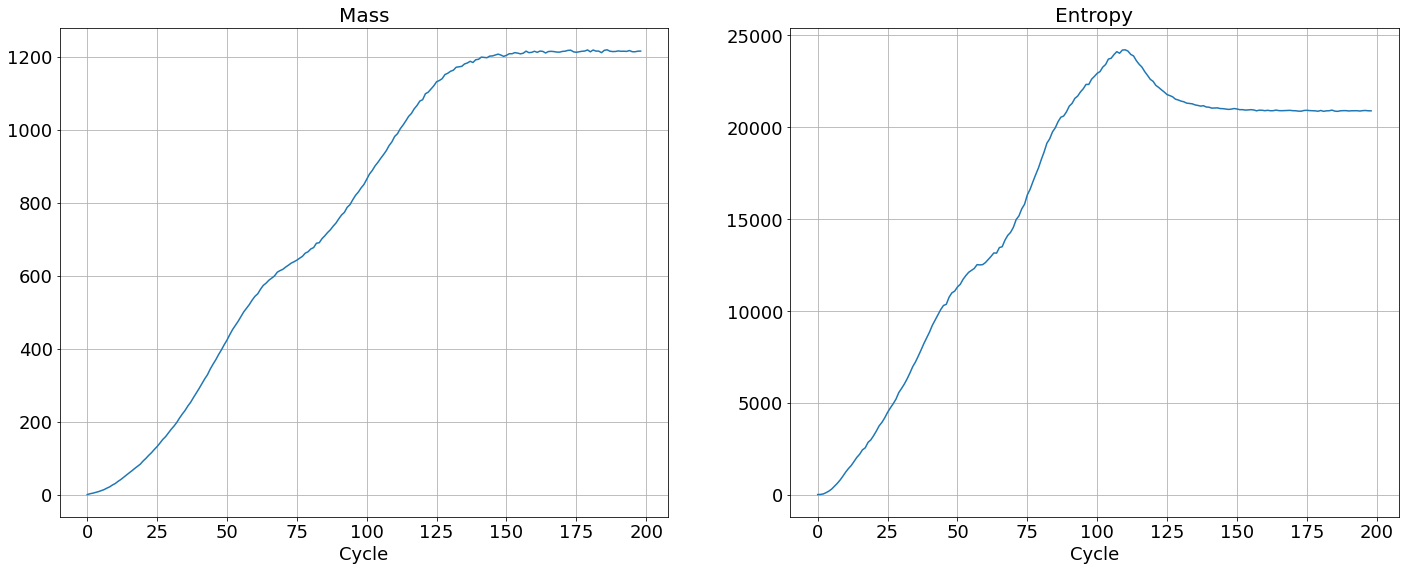} }}%
\caption{Diffusion process in \textbf{SCCGoL}p20L100b100 (lattice size 100 by 100, 20\% probability of
spontaneous change of cell state from life to dead and reversely with preservation of standard rules in Conway’s Game of Life as also given in Figure \ref{fig:life_expectancy}) case of system of two weekly interconnected chambers by means of two small holes in barrier. Two stages of diffusion can be spotted in mass and entropy dynamics that has two consecutive processes: full diffusion in the left chamber is leading to full diffusion in the right chamber. Further details of thermodynamical parameters space dependence evolution with time are given by Figure \ref{fig:characteristics_one_barrier}.}
\label{fig:one_barrier}
\end{figure}
We consider a single cellular automata (single automata seed) placed in empty chamber with impenetrable walls with two small holes that link it to another empty chamber as depicted in the Figure \ref{fig:one_barrier}a. We observe a diffusion of cellular automata with simulation time that consists of two main processes: creation and diffusion of cellular automata in the first chamber and diffusion of cellular automata from the first chamber into second chamber accompanied with creation new automata in the second chamber. Those two consequent processes are accompanied by effective slowdown in diffusion that is seen in left part of Figure \ref{fig:one_barrier}c. At the same time we observe slowdown in entropy increase as in an accordance to the right part of Figure \ref{fig:one_barrier}c. Once mass saturation was obtained in the simulation we observe maximization and small drop in entropy that later stabilizes and saturates, as in the right part of Figure \ref{fig:one_barrier}c.
\begin{figure}
\centering
\subfloat[Mass at t=5\centering]{{\includegraphics[width=.3\linewidth]{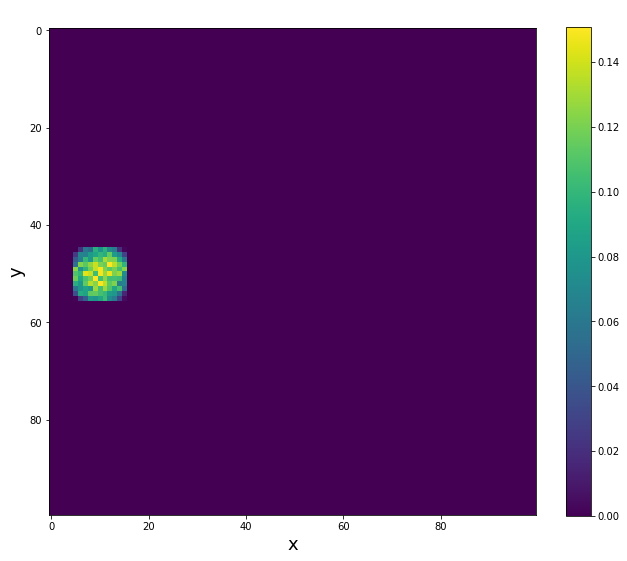} }}%
\subfloat[Mass at t=70\centering]{{\includegraphics[width=.3\linewidth]{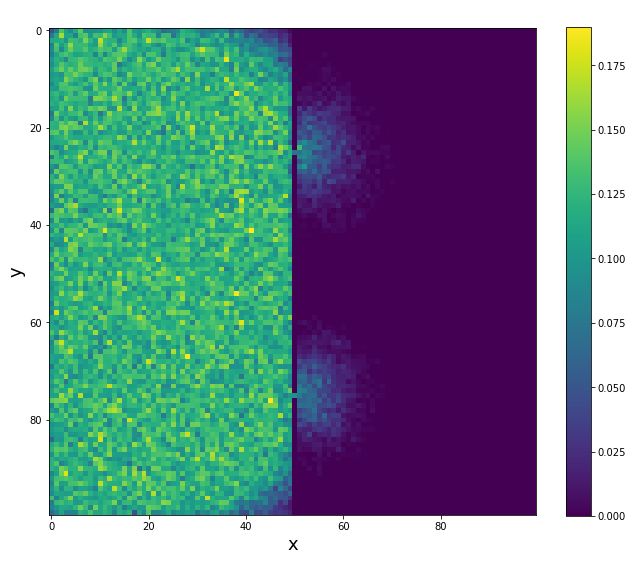} }}%
\subfloat[Mass at t=100\centering]{{\includegraphics[width=.3\linewidth]{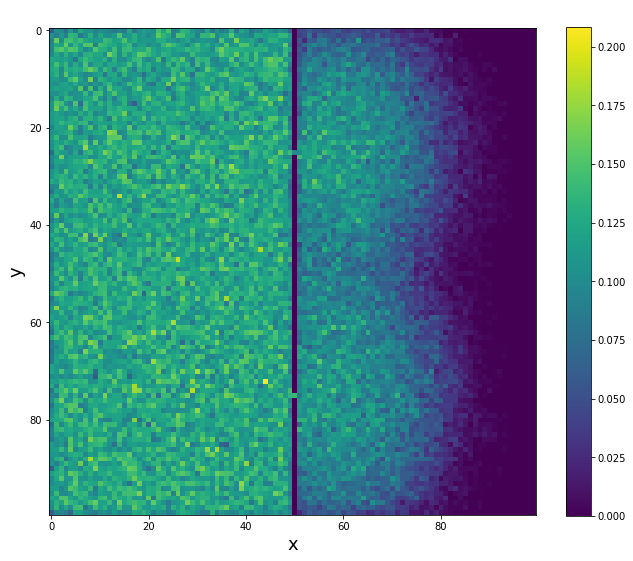} }}%
\qquad
\subfloat[Entropy at t=5\centering]{{\includegraphics[width=.3\linewidth]{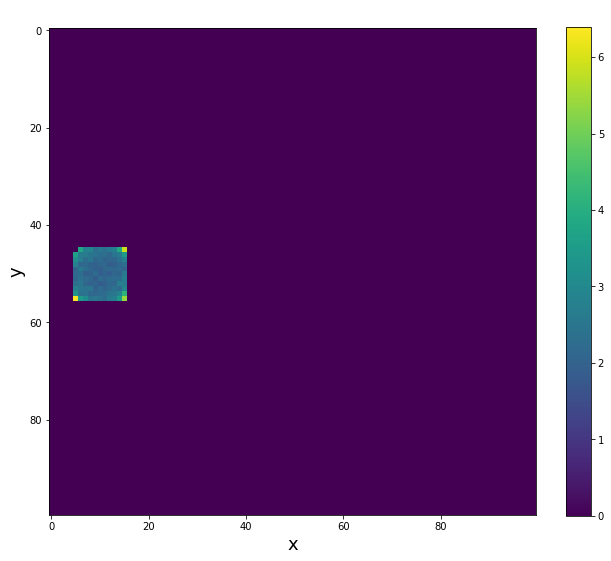} }}%
\subfloat[Entropy at t=70\centering]{{\includegraphics[width=.3\linewidth]{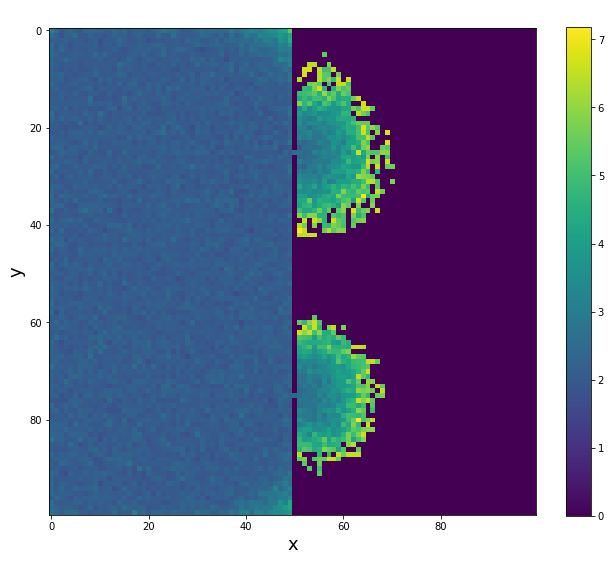} }}%
\subfloat[Entropy at t=100\centering]{{\includegraphics[width=.3\linewidth]{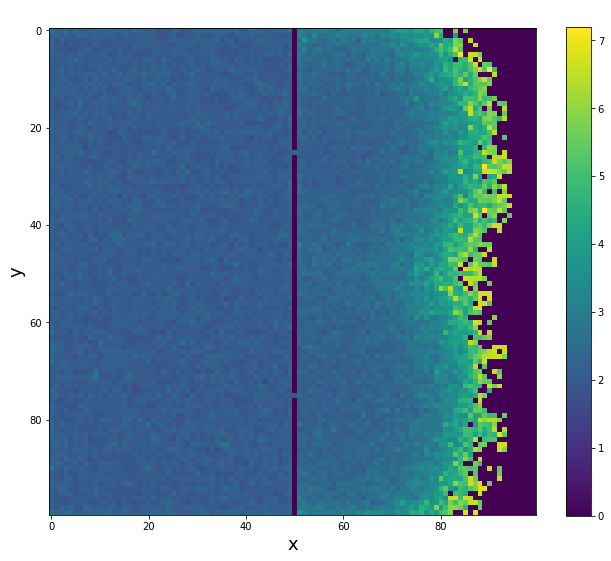} }}%
\qquad
\subfloat[T(x,y,t=5)\centering]{{\includegraphics[width=.3 \linewidth]{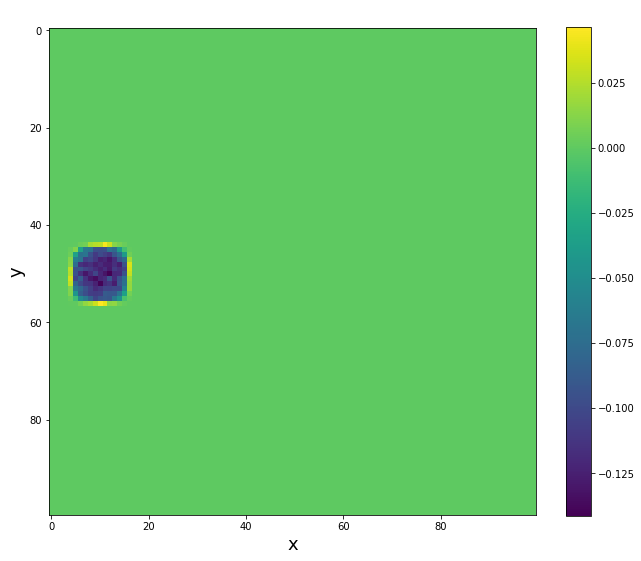} }}%
\subfloat[T(x,y,t=70)\centering]{{\includegraphics[width=.3 \linewidth]{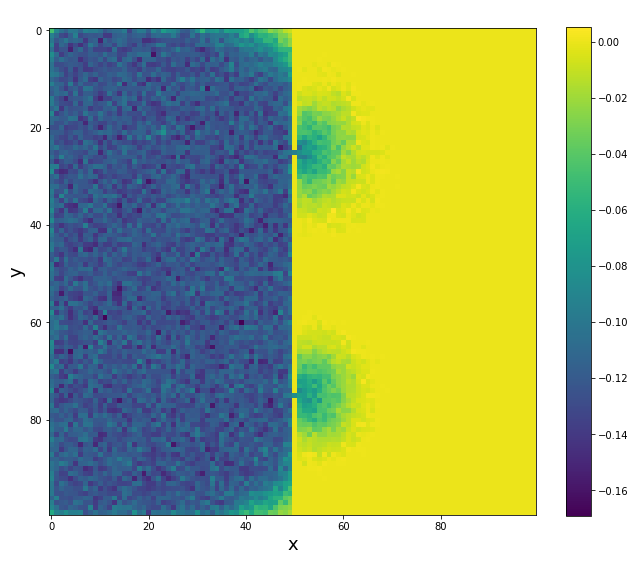} }}%
\subfloat[T(x,y,t=100)\centering]{{\includegraphics[width=.3 \linewidth]{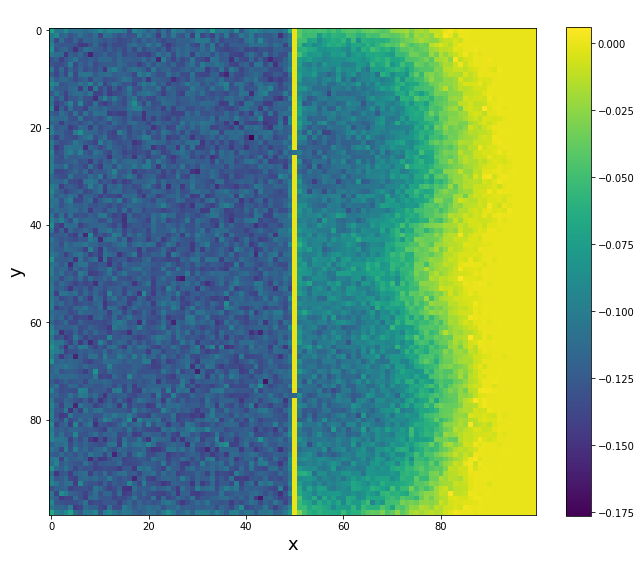} }}%
\qquad
\subfloat[$\frac{dm}{dt}$ with time\centering]{{\includegraphics[height=.22 \linewidth]{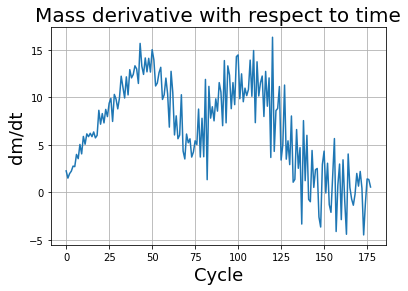} }}%
\subfloat[$\frac{dS}{dt}$ with time\centering]{{\includegraphics[height=.22 \linewidth]{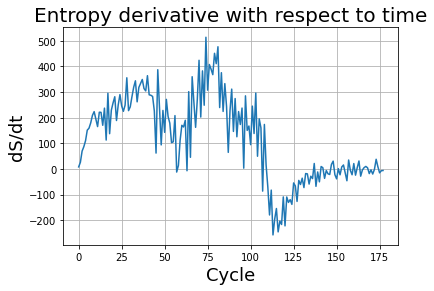} }}%
\subfloat[Temperature with time\centering]{{\includegraphics[height=.22 \linewidth]{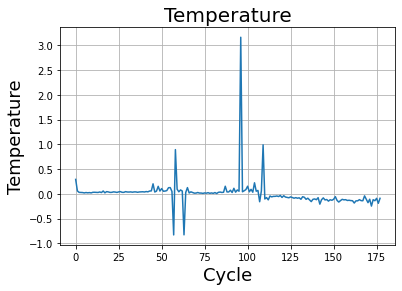} }}%
\caption{Dynamics of thermodynamical parameters with simulation time in \textbf{SCCGoL}p20L100b100 with two weekly interconnected chambers by two small holes in barrier that were initially depicted in Figure \ref{fig:one_barrier}.}
\label{fig:characteristics_one_barrier}
\end{figure}
Entropy unlike mass is characterized by a large variation of values in the middle of the population and at the edges of the cellular automata population. In the situation with barrier (case of Figure \ref{fig:characteristics_one_barrier}e) this is particularly evident after the cells pass through the gaps.
This particular process is due to the logarithm function dependence of Von Neumann entropy, which tends to negative infinity for arguments going to zero from the right. Observed criteria of an equilibrium is fact that mass and entropy have steady values and that temperature have negative value in thermodynamical equilibrium.
Almost always before the equilibrium moment is achieved, time derivatives of mass and entropy have positive values, and still the temperature is positive. From the moment the equilibrium is achieved, we are dealing with small fluctuations of entropy and temperature. We observe the correlation stating that slight increase in mass results in slight decrease in entropy and vice versa.
In an analogical way to the system with one barrier, simulations were conducted on the system with two barriers and the initial structure as depicted in Figure \ref{fig:two_barriers}a.
\begin{figure}
\centering
\subfloat[\centering]{{\includegraphics[width=.3 \linewidth]{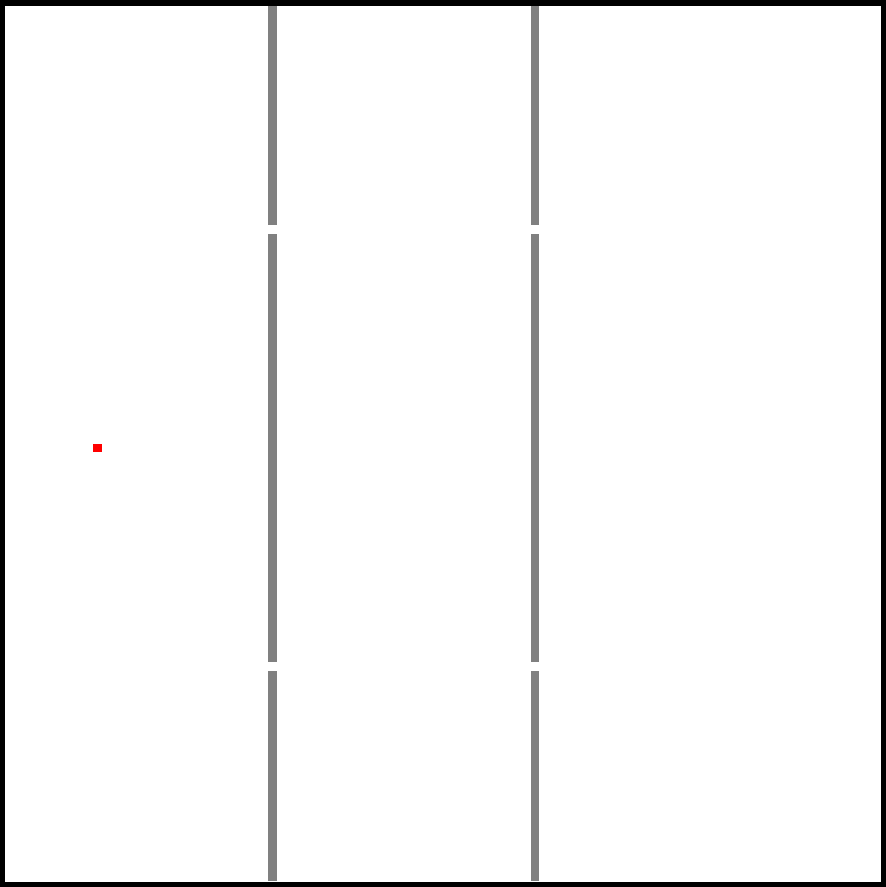} }}%
\qquad\qquad
\subfloat[\centering]{{\includegraphics[width=.3 \linewidth]{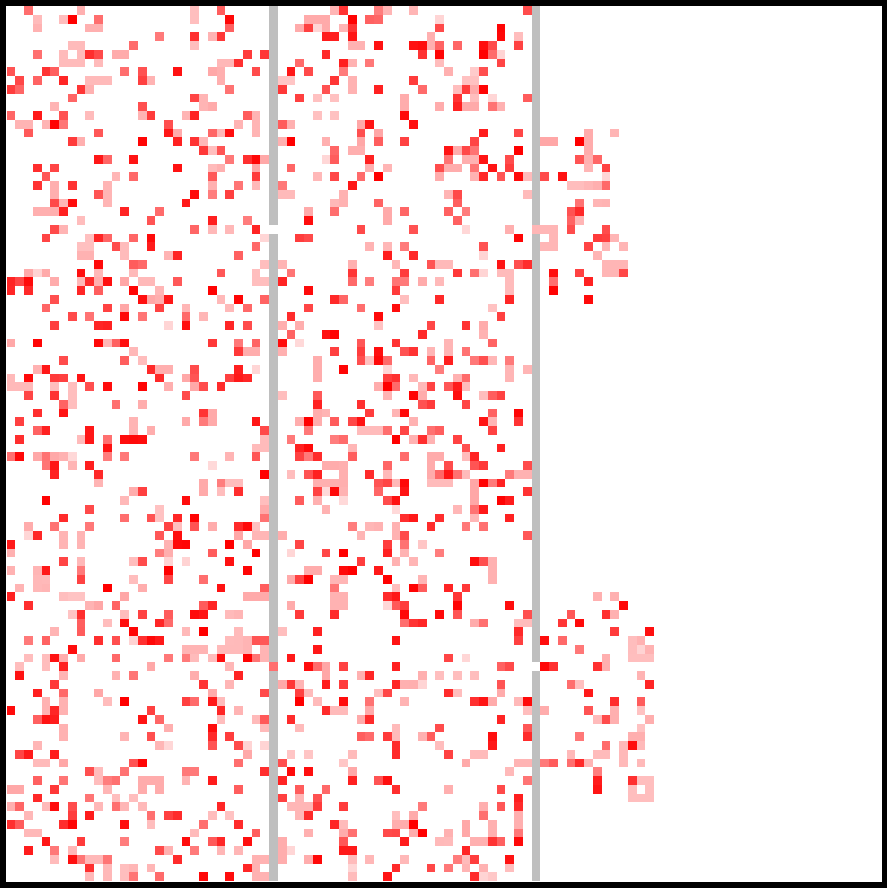} }}%
\qquad
\subfloat[\centering]{{\includegraphics[height=.3 \linewidth]{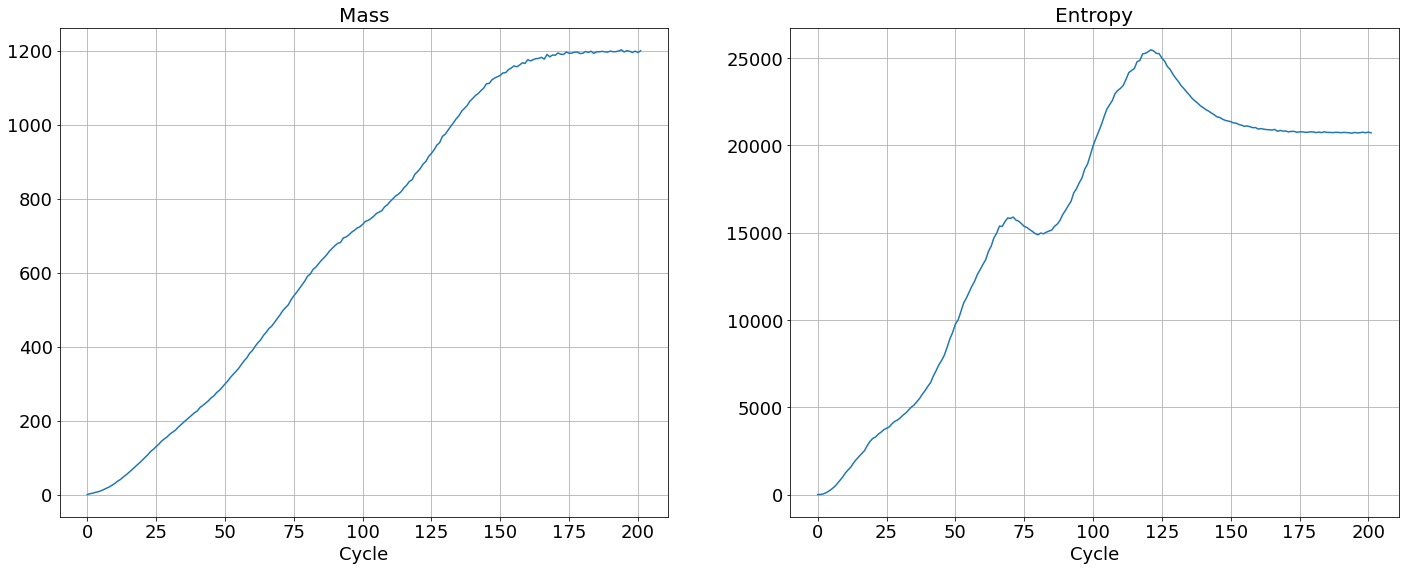} }}%
\caption{Dynamics of diffusion process for a system \textbf{SCCGoL}p20L100b100 with two barriers with two small holes in each barrier (generalization of situation from Figure \ref{fig:characteristics_one_barrier}) creating three weakly interconnected chambers perturbed by mutual interactions mediated by holes. Monotonicity in increase of entropy is twice shortly interrupted by small decline, what is associated with automata cells "colliding" with barriers and experiencing short lasting slowing down in its propagation. Details on space depended thermodynamical parameter evolution with time are given by Figure \ref{fig:characteristics_two_barriers}.}
\label{fig:two_barriers}
\end{figure}
We consider a single cellular automata placed in empty chamber with impenetrable walls with two small holes that link it to the second empty chamber, which is connected to a third empty chamber by impenetrable walls with two small holes as depicted in the Figure \ref{fig:two_barriers}a.
We observe a diffusion of cellular automata with simulation time that consists of three main processes: creation and diffusion of cellular automata in the first chamber, diffusion of cellular automata from the first chamber into second chamber accompanied with creation new automata in the second chamber and diffusion of cellular automata from the second chamber into third chamber accompanied with creation new automata in the third chamber. Twice cells try to reach impenetrable barriers, we observe small drops in entropy. After the second time entropy stabilizes and saturates, as in the right part of Figure \ref{fig:two_barriers}c.
\begin{figure}
\centering
\subfloat[Mass at t=30\centering]{{\includegraphics[width=.3\linewidth]{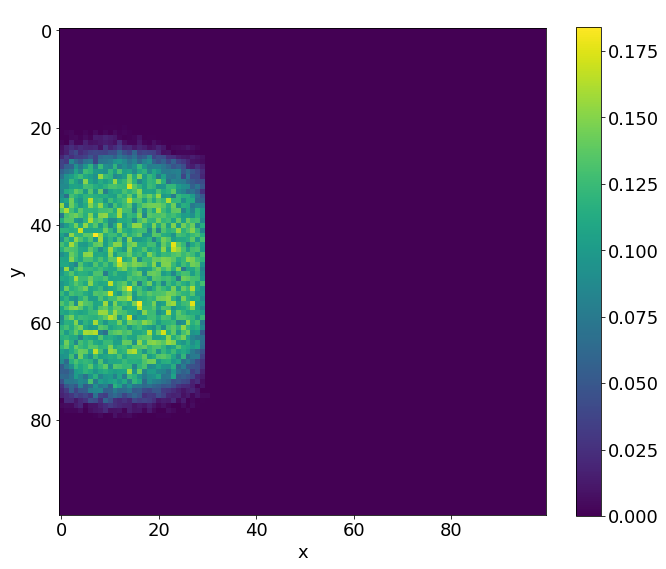} }}%
\subfloat[Mass at t=70\centering]{{\includegraphics[width=.3\linewidth]{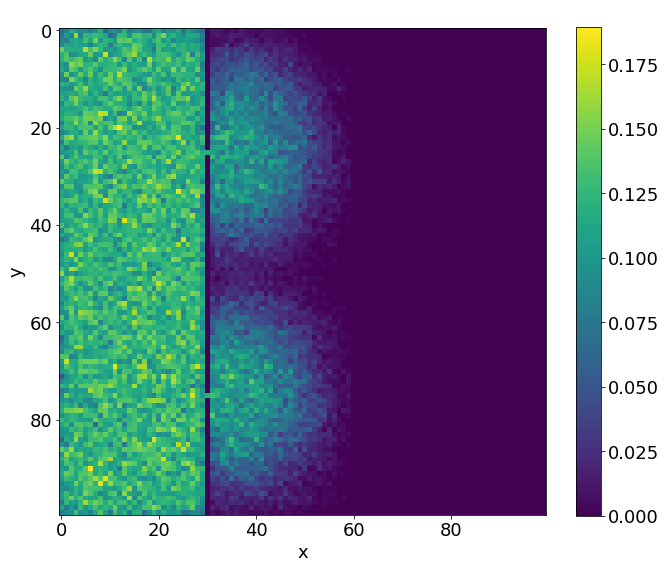} }}%
\subfloat[Mass at t=110\centering]{{\includegraphics[width=.3\linewidth]{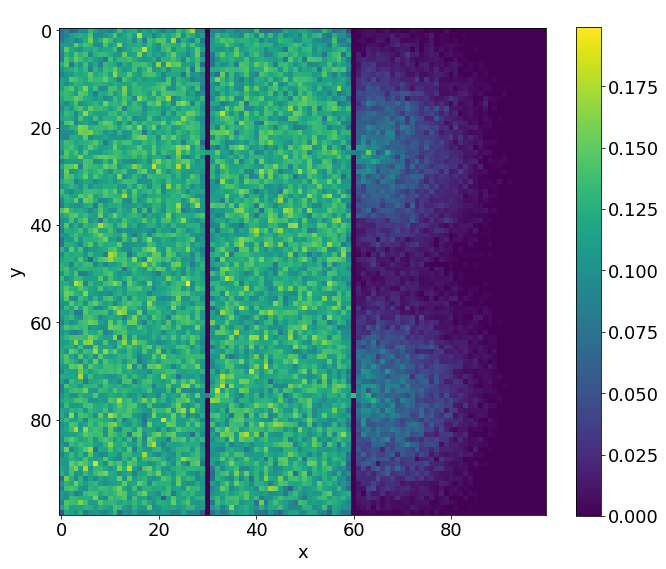} }}%
\qquad
\subfloat[Entropy at t=30\centering]{{\includegraphics[width=.3\linewidth]{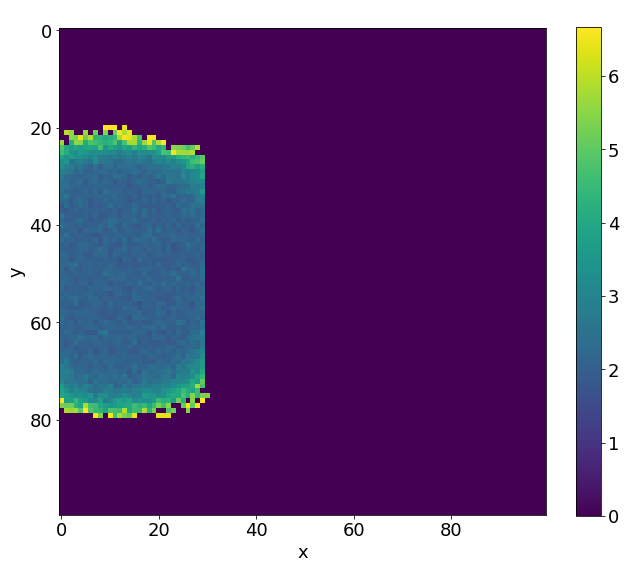} }}%
\subfloat[Entropy at t=70\centering]{{\includegraphics[width=.3\linewidth]{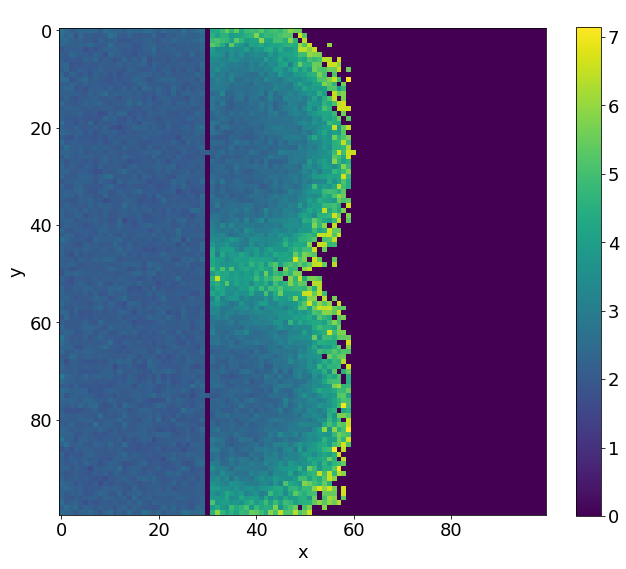} }}%
\subfloat[Entropy at t=110\centering]{{\includegraphics[width=.3\linewidth]{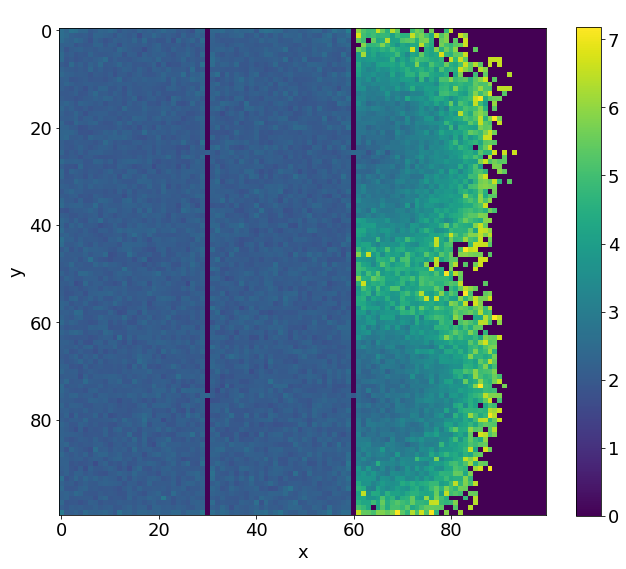} }}%
\qquad
\subfloat[T(x,y,t=30)\centering]{{\includegraphics[width=.3 \linewidth]{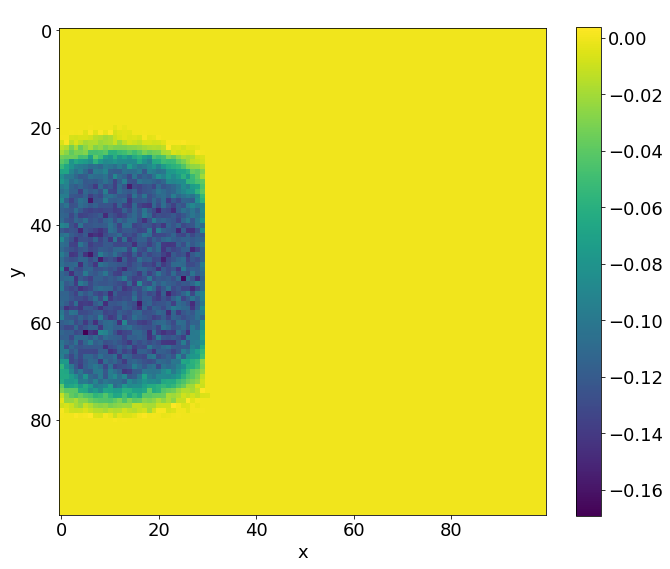} }}%
\subfloat[T(x,y,t=70)\centering]{{\includegraphics[width=.3 \linewidth]{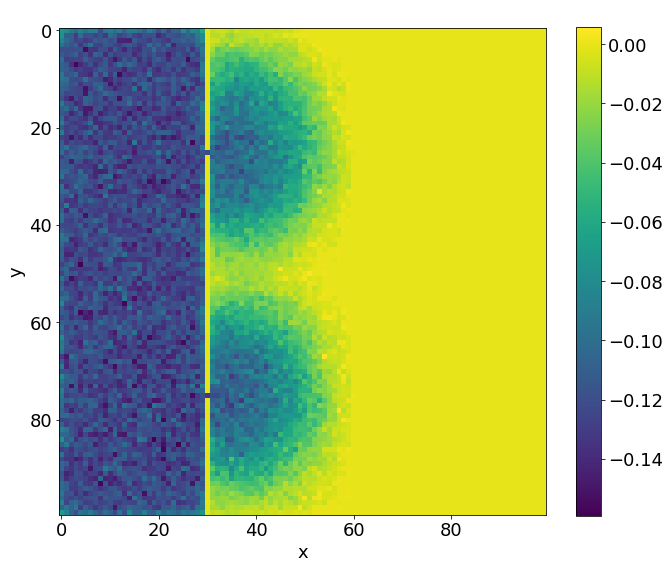} }}%
\subfloat[T(x,y,t=110)\centering]{{\includegraphics[width=.3 \linewidth]{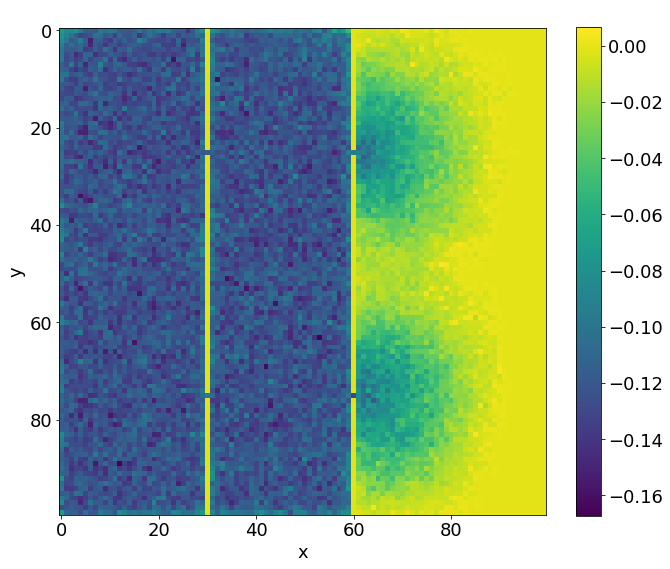} }}%
\qquad
\subfloat[$\frac{dm}{dt}$ with time\centering]{{\includegraphics[height=.215 \linewidth]{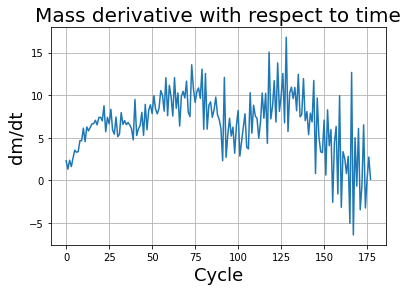} }}%
\subfloat[$\frac{dS}{dt}$ with time\centering]{{\includegraphics[height=.215 \linewidth]{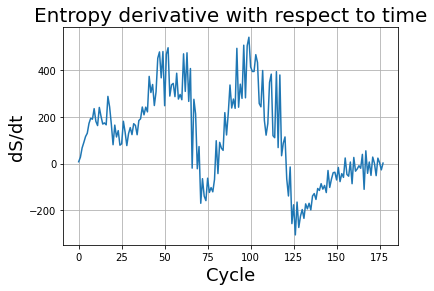} }}%
\subfloat[Temperature with time\centering]{{\includegraphics[height=.215 \linewidth]{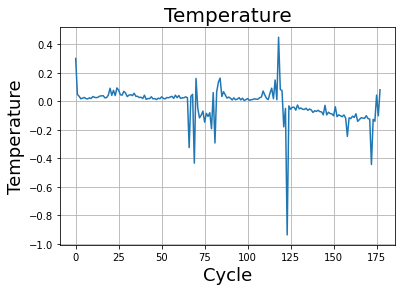} }}%
\caption{Space depended dynamics of thermodynamical parameters with simulation time in \textbf{SCCGoL}p20L100b100 with three weekly interconnected chambers by four small holes also depicted in Figure \ref{fig:two_barriers}.}
\label{fig:characteristics_two_barriers}
\end{figure}
In a system with two barriers, it lasts longer for mass and entropy to reach equilibrium than in the case of a system with only one barrier. As depicted in Figure \ref{fig:characteristics_two_barriers}k, due to the cells approaching the barriers and losing the extra entropy at the edges of the population, we observe significant fluctuations in the time derivative of the entropy. This results in large peaks seen in the Figure \ref{fig:characteristics_two_barriers}l.
\section{Numerical study of two species cellular automata in perturbative interaction by narrow constriction}
\begin{figure}
\centering
\subfloat[\centering]{{\includegraphics[width=.3 \linewidth]{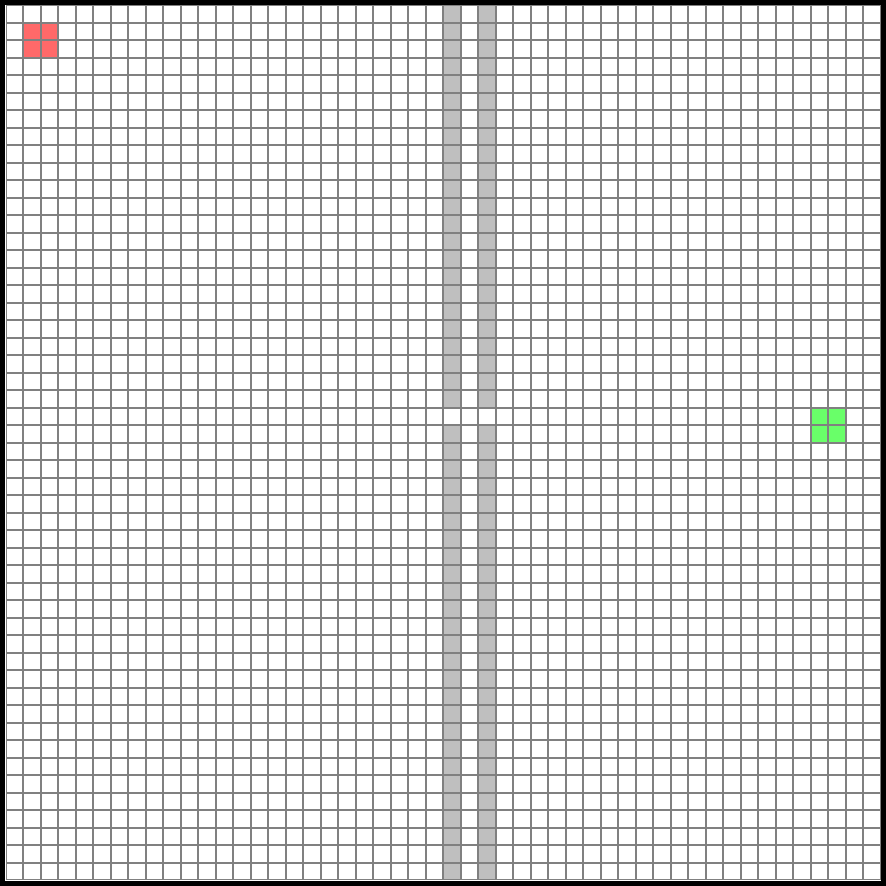} }}%
\qquad\qquad
\subfloat[\centering]{{\includegraphics[width=.3 \linewidth]{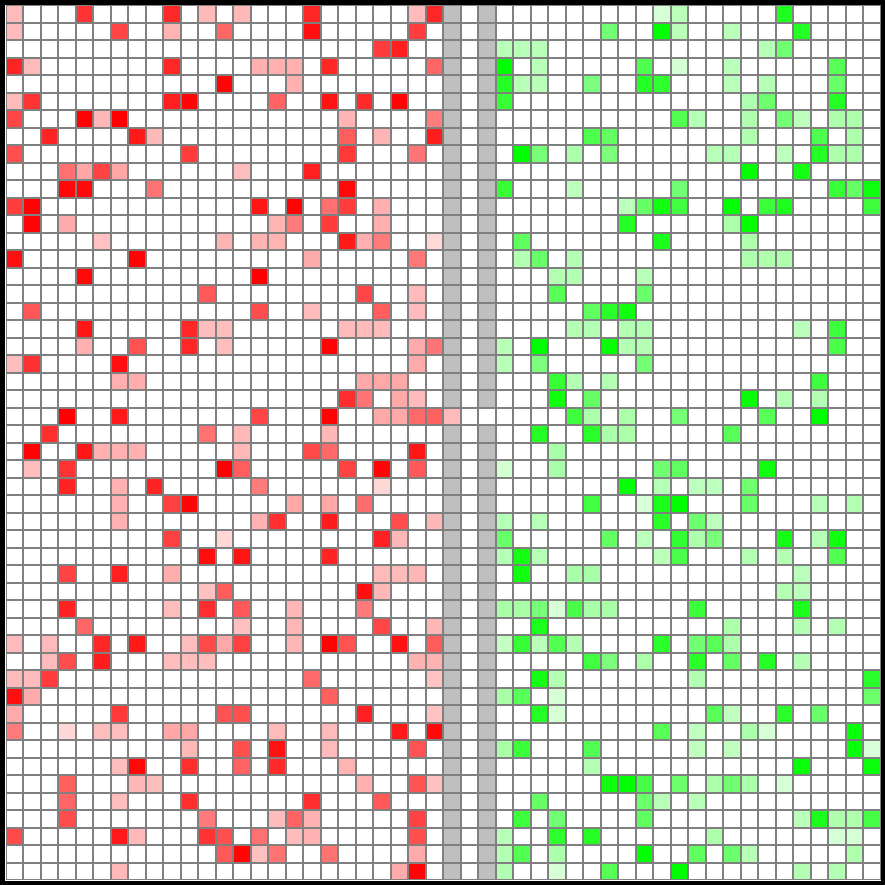} }}%
\qquad
\subfloat[\centering]{{\includegraphics[height=.3 \linewidth]{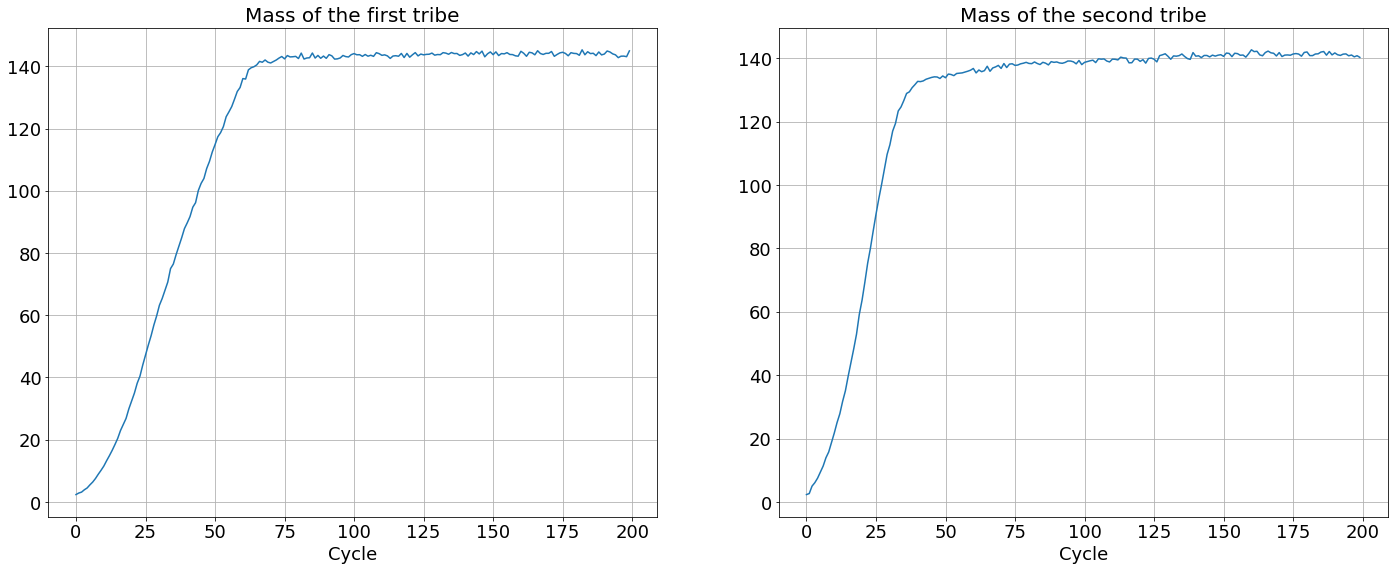} }}%
\qquad
\subfloat[\centering]{{\includegraphics[height=.3 \linewidth]{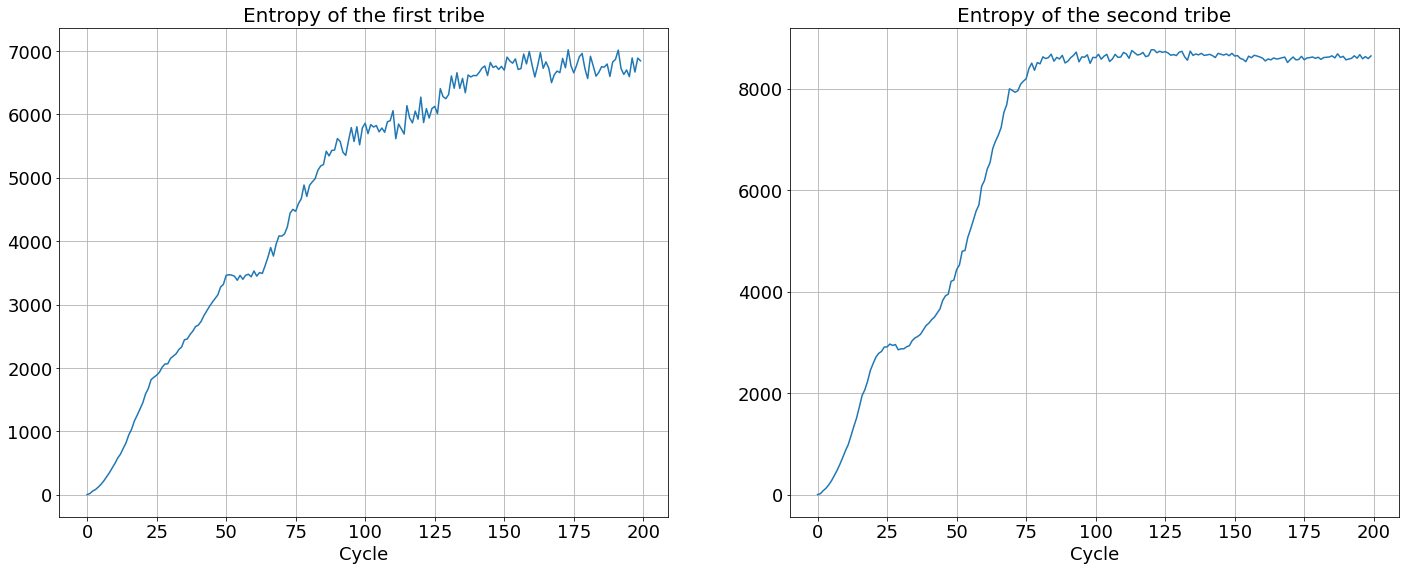} }}%
\qquad
\subfloat[Mass at t=3\centering]{{\includegraphics[width=.3\linewidth]{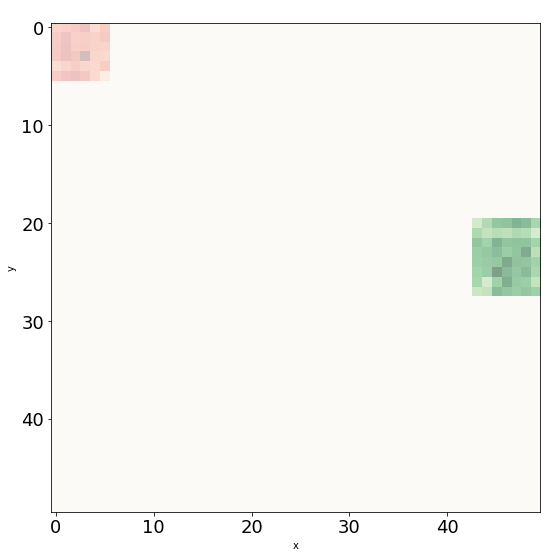} }}%
\subfloat[Mass at t=34\centering]{{\includegraphics[width=.3\linewidth]{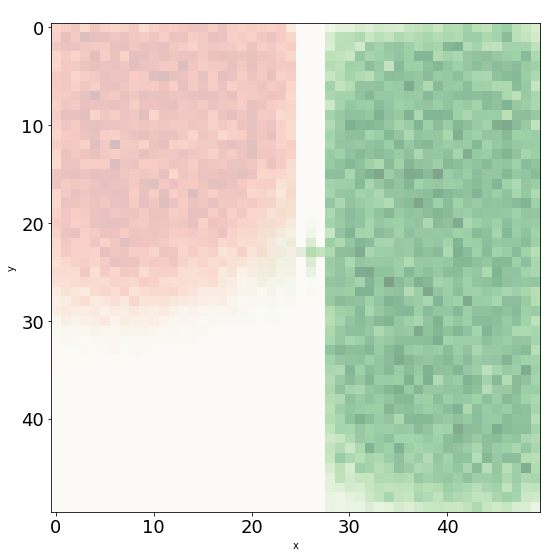} }}%
\subfloat[Mass at t=200\centering]{{\includegraphics[width=.3\linewidth]{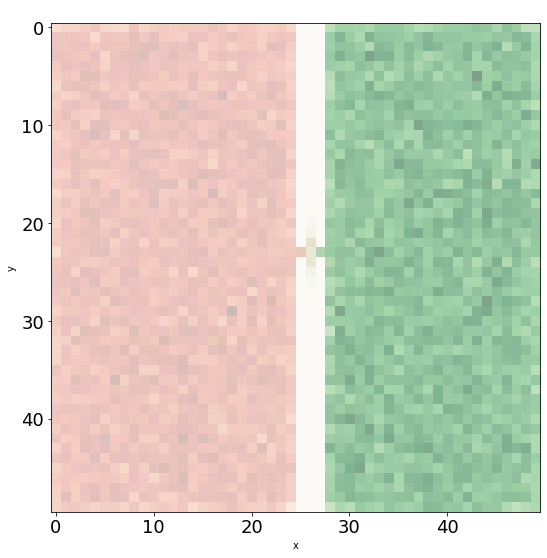} }}%
\caption{Diffusion process in case of system with two cellular automata tribes weekly interacting with each other via a small hole in double barrier as depicted in (a), where initial configuration is presented. After long thermodynamic equilibrium is achieved as given by (b) so two cellular automata tribes coexist in two different geometrical domains effectively geographically separated. In case of both tribes mass and entropy saturates having tendency to oscillate in thermodynamical equilibrium.}
\label{fig:two_tribes}
\end{figure}

Further simulations for the case of two species cellular automata were carried out for a system divided into two reservoirs separated by two impenetrable barriers with one small hole (case of Figure \ref{fig:two_tribes}a) that implies perturbative interaction between tribes.
In the left upper corner of the first part of the system there have been located cells of the first cellular automata tribe. At a closer distance to the hole in impenetrable wall, but in the second right reservoir there have been located cells of the second cellular automata tribe. As depicted in Figure \ref{fig:two_tribes}c, we observe similar final masses and their dynamics in case of both tribes, but in accordance to Figure \ref{fig:two_tribes}d, different entropy dynamics.
Very last is due to the distance of the cellular automata tribes from the small hole in impenetrable wall: a tribe located further from that hole needs more time to propagate and occupy its natural neighborhood and first left chamber of the system, and thus this tribe has a lower probability of taking over the territory of the other tribe. However the noticeable fact is that mass and entropy of both tribes achieves equilibrium and finally tribes end up in bit same geometrical and dynamical situation.
As depicted in Figure \ref{fig:two_tribes}f, the right tribe closer to the small hole occupies its nearest neighborhood territory more quickly, resulting in an attempt to occupy a rival tribe territory.
\begin{figure}
\centering
\subfloat[$\frac{dm}{dt}$ with time for first and second automata tribe\centering]{{\includegraphics[height=.3\linewidth]{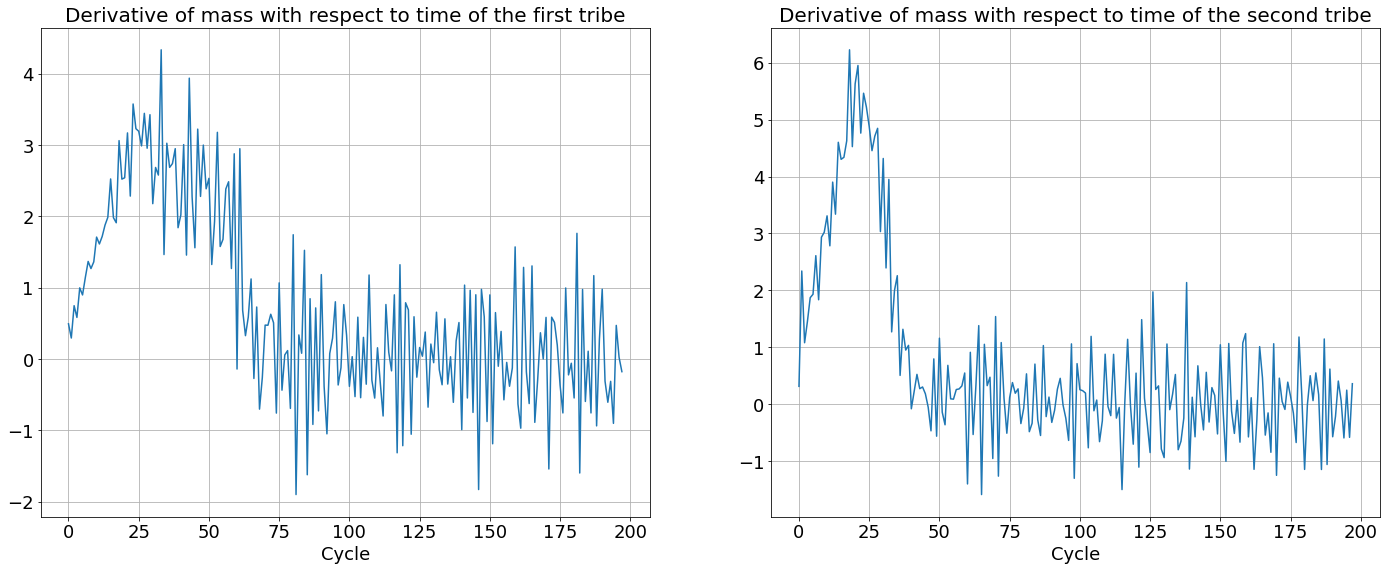} }}%
\qquad
\subfloat[$\frac{dS}{dt}$ with time for first and second automata tribe\centering]{{\includegraphics[height=.3\linewidth]{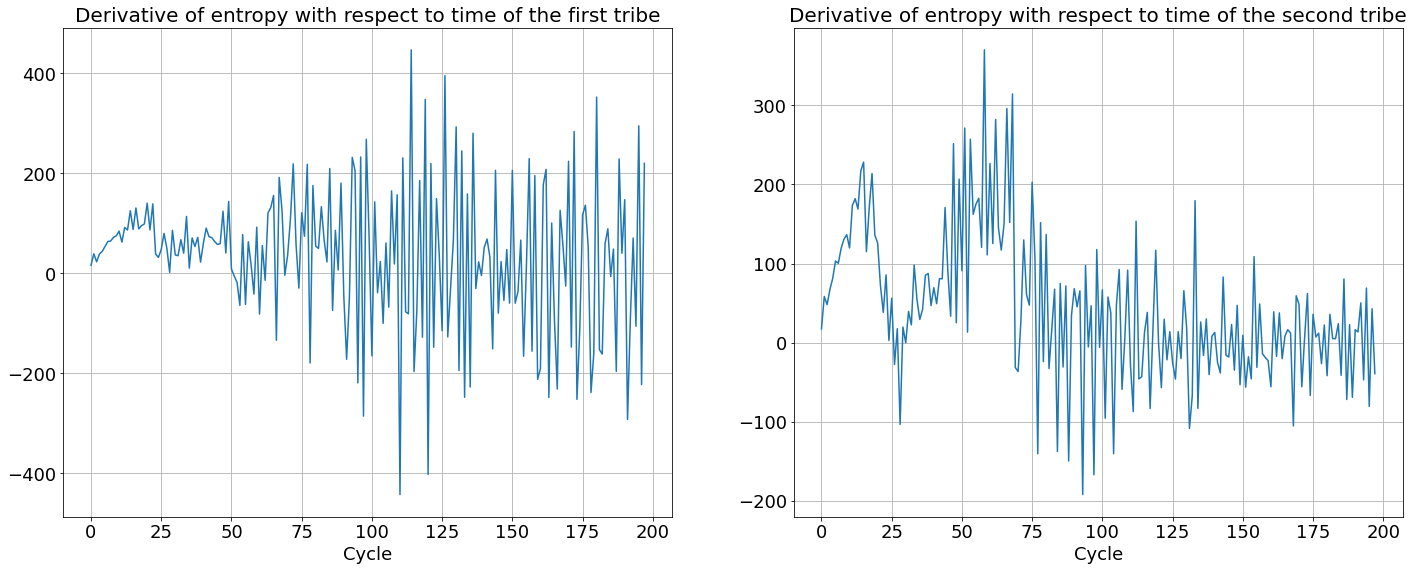} }}%
\qquad
\subfloat[Temperature with time for first and second automata tribe\centering]{{\includegraphics[height=.3\linewidth]{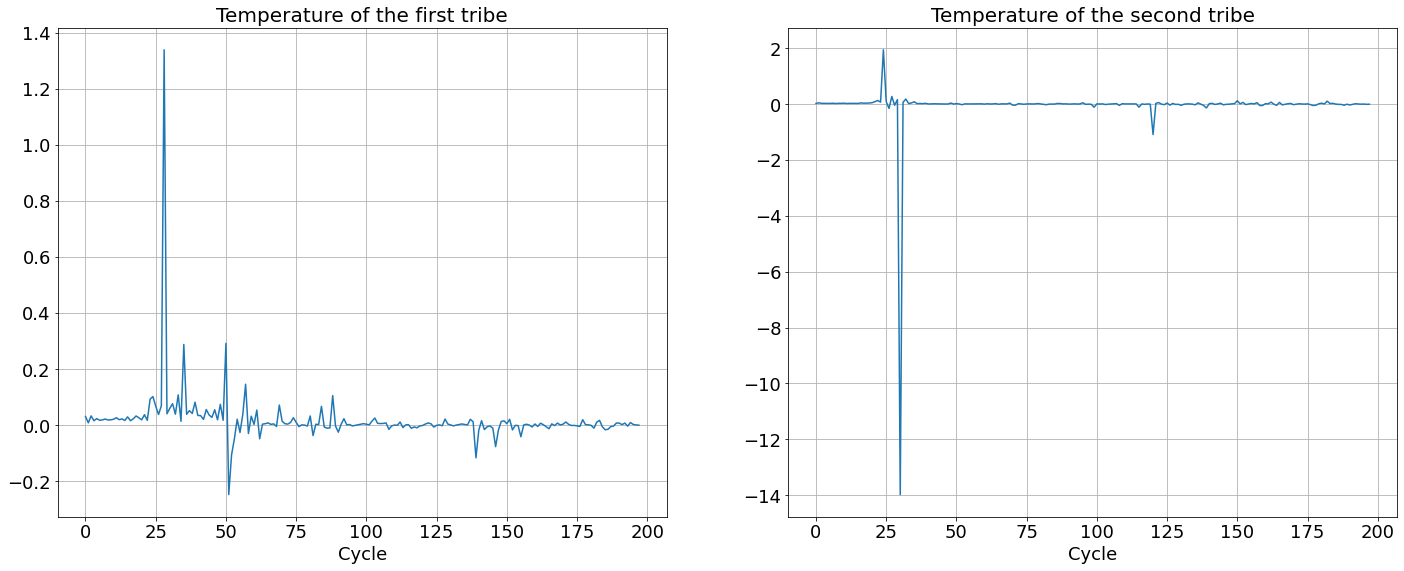} }}%
\caption{Dynamics of thermodynamical variables for the case of two competing cellular automata tribes (first tribe is placed on the left and second tribe is placed on the right).}
\label{fig:temperature_of_two_tribes}
\end{figure}
As depicted in Figure \ref{fig:temperature_of_two_tribes} we observe large oscillations in the time derivatives of mass and entropy of both cellular automata tribes. In contrast to previously conducted simulations, the temperature of the system after reaching equilibrium is not only characterized by negative values. Tribes existential competition is the reason of occurrence of both positive and negative temperatures.
\section{Conclusions and future perspectives}
Cellular automata can simulate many complex physical phenomena using the power of simple rules as it was shown in the case of cellular automata diffusion dynamics confirmed by various simulations for one and many automata species. Certain type of automata Darwinism was spotted by studying 4 automata species dynamics as given by Fig.\ref{fig:species2}.
The SCCGoL dynamics study provides strong evidence that despite the fact that the principle of conservation of mass is not fulfilled, since we have creationism and annihilation of automata, the entropy and temperature comes to equilibrium.
In conducted various simulations of Stochastic Conway's Game of Life dynamics we report transition from positive to negative values of temperatures and we are aware that there is maximum level of mass and energy density allowed for cellular automata, since otherwise they would die due to overpopulation.
The fact that the temperature can be negative is known in condensed matter physics, but with assumption that the energy is top-bounded. In most "normal" situations this is impossible, but in a rare cases in solid state physics approximately it can be achieved by inverting the population state. Obviously there is a such limitation on top-bounded energy value (mass density value) in Stochastic Conway's Game of Life. Therefore, it is still consistent with thermodynamics methodology, as it was pointed by professor Adam Bednorz (Faculty of Physics, University of Warsaw).
\\\\
Following conclusions were derived basing on conducted simulations and described methodology:
\begin{enumerate}
    \item Identification of thermodynamically defined temperature as proper measure of system evolution
with '-' sign (case of Figures \ref{fig:characteristics_no_barriers_1000}, \ref{fig:characteristics_one_barrier}, \ref{fig:characteristics_two_barriers}).
    \item Identification of mass as effective energy of system (in first approximation) (case of Figures \ref{fig:mass_entropy_no_barriers_1000}, \ref{fig:one_barrier}, \ref{fig:two_barriers}, \ref{fig:two_tribes}).
    \item Identification of Shannon Entropy as effective system entropy (in first approximation) (case of Figures \ref{fig:mass_entropy_no_barriers_1000}, \ref{fig:one_barrier}, \ref{fig:two_barriers}, \ref{fig:two_tribes}).
    \item Generalization of Stochastic Conway Game of Life of $N$ tribes
(approximated analogy to $N$-body Quantum Physics can be conceptionally drawn) as depicted in Figures \ref{fig:species2}, \ref{fig:two_tribes}.
    \item Confirmation validity of second law of thermodynamics in SCCGoL (entropy maximises and saturates, case of Figures \ref{fig:mass_entropy_no_barriers_1000}, \ref{fig:one_barrier}, \ref{fig:two_barriers}).
    \item Identification of short lasting Shannon entropy peak that later minimizes and saturates in SGoL (case of Figures \ref{fig:mass_entropy_no_barriers_1000}, \ref{fig:one_barrier}). Monotonicity in increase of entropy is twice shortly interrupted by small decline, what is associated with automata cells "colliding" with barriers and experiencing short lasting slowing down in its propagation (case of Figure \ref{fig:two_barriers}).
\end{enumerate}
Conducted analysis of Stochastic Game of Life allows to treat such system as mathematical object well described by methodology of classical statistical physics. Obtained numerical results by various simulations suggest that we shall introduce another definition of temperature in Stochastic Conway's Game of Life system by adding 'minus' sign to temperature known in statistical physics, so we obtain the following formula:
\begin{equation}
Temperature_{Conway-Pomorski-Kotula}=-\frac{dE}{dS}
\end{equation}
$$Temperature_{Statistical Physics}=+\frac{dE}{dS}$$
Having such a definition of Conway-Pomorski-Kotula temperature we can use tools of statistical physics in Stochastic Game of Life preserving most classical physics thermodynamical intuition about various situations we can come across. There are well-known inherent analogous between classical statistical physics \cite{cite:9}\cite{cite:10}\cite{cite:11}\cite{cite:12}\cite{cite:13} and quantum mechanics \cite{cite:14}.
Therefore further research perspectives in study of Stochastic Classical Conway's Game of Life assume the usage of quantum mechanics being able to simulate classical statistical physics (as expressed by epidemic model or stochastic finite-state machine) as explicitly represented by tight-binding model \cite{cite:7}\cite{cite:8} or Schroedinger model directly proposing structures implemented in semiconductor single-electron devices. Noticeable various obtained map of probability of Stochastic Conway's Game of Life can be parameterized by non-linear Schrodinger equation and especially by Ginzburg-Landau model \cite{cite:6}, what can be the base for quantization of Stochastic Classical Conway's Game of Life.
\section{Acknowledgment}
We would like to express our acknowledgment to professor Adam Bednorz (University of Warsaw), Adam Chochla (Cracow University of Technology) and to doctor Lukasz Stepien (The Pedagogical University in Cracow). The consultations with them on manuscript has allowed to improve it. The Authors have no conflict of interests and equally contributed to this work with 50 percent of contribution on each side. First Author proposed methodological and conceptual framework for this work, while Second Author conducted all numerical simulations. The interpretations of obtained results is equally assigned to each Author.
%
%
%
\bibliographystyle{splncs04}
\bibliography{biblio}
\end{document}